\documentclass[a4paper,14pt]{article}
\usepackage[english]{babel}
\pdfoutput=1 % if your are submitting a pdflatex (i.e. if you have
\usepackage{jheppub} % for details on the use of the package, please
\usepackage[T1]{fontenc} % if needed

\usepackage[T1,T2A]{fontenc}
\usepackage[utf8x]{inputenc}

\usepackage{comment}

\usepackage{cases}

\newcommand{\bi}{\begin{itemize}}
    \newcommand{\ei}{\end{itemize}}
\newcommand{\bea}{\begin{eqnarray}}
    \newcommand{\eea}{\end{eqnarray}}
\newcommand{\bt}{\begin{tabular}}
    \newcommand{\et}{\end{tabular}}
\newcommand{\bc}{\begin{center}}
    \newcommand{\ec}{\end{center}}

\newcommand{\be}{\begin{equation}}
    \newcommand{\ee}{\end{equation}}
\newcommand{\ba}{\begin{array}}
    \newcommand{\ea}{\end{array}}

\newcommand{\lb}[1]{\label{#1}}

\def\bbox{{\,\lower0.9pt\vbox{\hrule \hbox{\vrule height 0.2 cm
                \hskip 0.2 cm \vrule height 0.2 cm}\hrule}\,}}
\newcommand{\dsl}{\pa \kern-0.5em /}

\makeatletter \@addtoreset{equation}{section} \makeatother

\def\slashchar#1{\setbox0=\hbox{$#1$}           % set a box for #1
    \dimen0=\wd0                                 % and get its size
    \setbox1=\hbox{/} \dimen1=\wd1               % get size of /
    \ifdim\dimen0>\dimen1                        % #1 is bigger
    \rlap{\hbox to \dimen0{\hfil/\hfil}}      % so center / in box
    #1                                        % and print #1
    \else                                        % / is bigger
    \rlap{\hbox to \dimen1{\hfil$#1$\hfil}}   % so center #1
    /                                         % and print /
    \fi}

\pdfoutput=1

\title{\boldmath  $\mathcal{N}=2$ AdS hypermultiplets in harmonic superspace}

\author[a,b]{Evgeny~Ivanov,}
\author[a,b]{Nikita~Zaigraev}

\affiliation[a]{Bogoliubov Laboratory of Theoretical Physics, JINR,\\141980 Dubna, Moscow region, Russia}
\affiliation[b]{Moscow Institute of Physics and Technology,\\ 141700 Dolgoprudny, Moscow region, Russia}

\emailAdd{eivanov@theor.jinr.ru}
\emailAdd{nikita.zaigraev@phystech.edu}

\abstract{We present the harmonic superspace formulation of $\mathcal{N}=2$ hypermultiplet in AdS$_4$ background, starting from the proper realization of $4D, \mathcal{N}=2$ superconformal group $SU(2,2|2)$ on the analytic
subspace coordinates. The key observation is that $\mathcal{N}=2$ AdS$_4$ supergroup $OSp(2|4)$ can be embedded as a subgroup in the superconformal group through introducing a constant symmetric matrix $c^{(ij)}$ and
identifying the AdS supercharge as $\Psi^i_\alpha = Q^i_\alpha + c^{ik} S_{k\alpha}$, with $Q$ and $S$ being generators of the standard and conformal $4D, {\cal N}=2$ supersymmetries. Respectively, the AdS cosmological
constant is given by the square of $c^{(ij)}$, $\Lambda = -12 c^{ij}c_{ij}$. We construct the $OSp(2|4)$ invariant hypermultiplet mass term by adding, to the coordinate AdS transformations, a piece realized as an extra
$SO(2)$ rotation of the hypermultiplet superfield. It is analogous to the central charge $x^5$ transformation of  flat $\mathcal{N}=2$ supersymmetry and turns into the latter in the super Minkowski limit. As another new
result, we explicitly construct the superfield Weyl transformation to the $OSp(2|4)$ invariant AdS integration measure over the analytic superspace, which provides, in particular, a basis for unconstrained superfield
formulations of the AdS$_4$-deformed $\mathcal{N}=2$ hyper K\"ahler sigma models. We find the proper redefinition of $\theta$ coordinates ensuring the AdS-covariant form of the analytic superfield component expansions. }

\arxivnumber{2509.01406}

\makeatletter
\gdef\@fpheader{}
\makeatother

\begin{document}

\maketitle
\flushbottom

%\newpage

\section{Introduction}\label{sec: intro}

The Harmonic Superspace (HSS) provides an appropriate arena for formulating and studying  theories with extended off-shell $\mathcal{N}=2$ Poincar\'e supersymmetry
and  $\mathcal{N}=2$ superconformal symmetry, including higher-spin
theories \cite{18,Galperin:1984av,  Buchbinder:2021ite, Galperin:1987ek, Galperin:1987em}. However, the full-fledged formulation of theories with
the explicit $\mathcal{N}=2$ AdS supersymmetry $\mathfrak{osp}(2|4)$ in HSS is still missing.

\smallskip

The present paper is intended to partially fill in this gap. Starting from the fundamental matter $\mathcal{N}=2$ multiplet, hypermultiplet, we construct
some superfield models which are invariant under $\mathcal{N}=2$ AdS supersymmetry properly realized in HSS.
As elaborated in \cite{Buchbinder:2022kzl, Buchbinder:2022vra, Buchbinder:2024pjm}
(for review see \cite{Buchbinder:2024xll, Buchbinder:2025ceg}), unconstrained analytic higher-spin prepotentials in the flat super-Minkowski background
can be naturally derived from gauging  rigid symmetries of the free hypermultiplet. We expect that this procedure admits a natural extension to the AdS$_4$  superspace,
so the present paper should be considered as  the first step towards construction of the complete  harmonic formulation of
$\mathcal{N}=2$ AdS higher-spin theories\footnote{Off-shell $4D, \mathcal{N}=2$ AdS higher-spin supermultiplets can be formulated in terms of
$\mathcal{N}=1$ superfields \cite{Kuzenko:1994dm, Gates:1996xs, Segal:1999qr, Gates:1996my}. However, the complete  off-shell formulations of such theories in ${\cal N}=2$
HSS were missed until now. We believe that such formulations not only will  manifest the beautiful geometric underlying
structure of these theories, but can also shed additional light on the problem of constructing consistent nonlinear $\mathcal{N}=2$ higher-spin supergravities.
In particular, we expect that the fundamental quantities of the latter are unconstrained prepotentials living on the AdS version of ${\cal N}=2$ harmonic analytic superspace.}.

\medskip

Our approach is based on the observation that the superalgebra $\mathfrak{osp}(2|4)$  can be embedded as a subalgebra in the ${\cal N}=2$ superconformal algebra $\mathfrak{su}(2|2,2)$
by making use of constant $SU(2)$ breaking triplet $c^{ik}$ \cite{Gates:1983nr, Bandos:2002nn}.
The analyticity-preserving realization of $\mathcal{N}=2$ superconformal algebra on the  HSS coordinates is well known \cite{18, Galperin:1985zv, Buchbinder:2024pjm},
so one can easily find the analogous realization of $\mathcal{N}=2$ AdS superalgebra. Being aware of such transformations, one can add terms that
break the superconformal invariance of the free massless hypermultiplet down to
$\mathfrak{osp}(2|4)$ invariance. Characteristic examples of such $\mathfrak{osp}(2|4)$ invariant systems are the hypermultiplet AdS mass term
%(both in superspace and in the ordinary AdS$_4$ space),
as well as some interacting sigma-model type superfield Lagrangians.
The crucial role in our approach is played by the super Weyl rescaling of the hypermultiplet superfield. It naturally brings the standard analytic integration measure which is manifestly
invariant under ${\cal N}=2$ Poincar\'e supersymmetry into the measure which is manifestly invariant under the AdS$_4$ superymmetry $\mathfrak{osp}(2|4)$.
%\medskip

The significant difference of the approach used in our paper from the approach of \cite{Kuzenko:2007aj, Kuzenko:2007vs, Kuzenko:2008qw} (as applied, e.g., to theories with $5D, \mathcal{N}=1$ AdS supersymmetry), consists in
the following. The second approach starts from the analysis of the superalgebra of the full set of covariant derivatives in the central basis of the AdS superspace one deals with. This approach is actually a generalization
of the one for $4D, \mathcal{N}=1$ AdS supersymmetry pioneered to the full extent in \cite{Ivanov:1980vb}. Our departure point is a realization of ${\cal N}=2$ superconformal supersymmetry in the analytic harmonic superspace
and on the unconstrained hypermultiplet $q^{+}$ superfields. Many peculiarities of these realizations  are inherited by the super AdS$_4$ harmonic formalism. For example, one of the most striking distinctions of our approach
is that the harmonic variables possess non-trivial $\mathfrak{osp}(2|4)$ transformation laws, while in the approach of  \cite{Kuzenko:2007aj} the harmonic variables (or their projective superspace avatars) are inert under
AdS supergroups. These two approaches are complementary to each other\footnote{Note that in \cite{Kuzenko:2008qw} the triplet constant analogous to $c^{ik}$ arises as the result of choosing a specific gauge for the
appropriate torsion component. More information about the possible origin of the constant $c^{ij}$ can be found in various works on supergravity, see, e.g.,  ref. \cite{deWit:1987sn} and references therein. } and we plan to
study details of their interplay elsewhere. We point out that our paper should be looked upon as a prerequisite to constructing the complete self-consistent superfield theory of higher spins in AdS$_4$ HSS. We also postpone
the detailed component analysis of the models considered here to further publications, focusing basically on their superfield structure.
%\medskip

\smallskip

The paper is organized as follows. In section \ref{sec 2} we describe $\mathcal{N}=2$ AdS supersymmetry as a subset of $\mathcal{N}=2$ superconformal symmetry.
In section \ref{sec: 21} we present  the explicit realization of $\mathcal{N}=2$ AdS supergroup
in harmonic superspace. Section \ref{sec 3} is devoted to the construction of the hypermultiplet mass term breaking ${\cal N}=2$ superconformal symmetry down to the AdS$_4$ ${\cal N}=2$
supersymmetry. In section \ref{sec 4} we present one of the main results of the paper: the explicit form of superfield Weyl rescaling leading to the
AdS analytic superspace integration measure. In Appendix \ref{eq: App A} we collect a number of useful notations, identities and  some important transformation laws. In the Appendix \ref{app B}
we present a general redefinition of Grassmann coordinates leading to the AdS covariant component fields in the superfield $\theta$-expansions.

%\newpage

\section{$\mathcal{N}=2$ AdS superalgebra as a subalgebra of $\mathfrak{su}(2,2|2)$ } \label{sec 2}

One of the important properties of the AdS (super)algebras is that they form subalgebras in the appropriate (super)conformal algebras (see, e.g., \cite{Bandos:2002nn}).
In $\mathcal{N}=2$
case, the non-trivial (anti)commutation relations of superconformal algebra $\mathfrak{su}(2,2|2)$ are given by\footnote{We use the following  conventions: $P_{\alpha\dot{\alpha}} := \frac{1}{2} \sigma_{\alpha\dot{\alpha}}^mP_m$, $K_{\alpha\dot{\alpha}} := \frac{1}{2} \sigma_{\alpha\dot{\alpha}}^mK_m$,
$L_{(\alpha\beta)}:= \frac{1}{2} \sigma^{mn}_{(\alpha\beta)}L_{mn}$, $\bar{L}_{(\dot{\alpha}\dot{\beta})} = - \frac{1}{2} \bar{\sigma}_{(\dot{\alpha}\dot{\beta})}^{mn} L_{mn}$.}:
\begin{subequations}\label{eq: N=2 SC}
         \begin{equation}\label{eq: N=2 SC a}
    \begin{split}
        &\{Q_\alpha^i, \bar Q_{\dot\alpha k}\} =
        4 \delta^i_k P_{\alpha\dot\alpha} ~, \quad \{S_{\alpha k}, \bar S_{\dot\alpha}^i\} =
        4\delta^i_k K_{\alpha\dot\alpha}~,
        \\
        & \{Q_\alpha^i, S^{\beta k}\} = 2\varepsilon^{ik}
        L_\alpha^\beta + 2i\varepsilon^{ik}\delta^\beta_\alpha (D + iR) -4i
        \delta^\beta_\alpha T^{(ik)}~,
   \end{split}
\end{equation}
    \begin{equation}
    \begin{split}
        & [D, Q] = \frac{i}{2}Q,
        \quad
        [D, \bar{Q}] = \frac{i}{2} \bar{Q},
        \quad
        [D, S] = - \frac{i}{2}S,
        \quad
        [D, \bar{S}] = -\frac{i}{2} \bar{S},
            \\
        & [R, Q] =- \frac{1}{2}Q,
        \quad
        [R, \bar{Q}] = \frac{1}{2} \bar{Q},
        \quad
        [R, S] =  \frac{1}{2}S,
        \quad
        [R, \bar{S}] = -\frac{1}{2} \bar{S},
   \end{split}
\end{equation}
\begin{equation}
\begin{split}
        & [P_{\alpha\dot{\alpha}}, K_{\beta\dot{\beta}}]
        =
        -i \varepsilon_{\alpha\beta}\varepsilon_{\dot\alpha\dot\beta} D
        +
        \frac{1}{2} \left( \varepsilon_{\alpha\beta} \bar{L}_{(\dot{\alpha}\dot{\beta})} + \varepsilon_{\dot\alpha\dot\beta} L_{\alpha\beta} \right),
  \end{split}
\end{equation}
\begin{equation}
\begin{split}
        & [K_{\alpha\dot{\alpha}}, Q^i_\beta] = \varepsilon_{\alpha\beta} \bar S^{ i}_{\dot\alpha}~,
        \quad [P_{\alpha\dot{\alpha}}, S_{ \beta i}] = \varepsilon_{\alpha\beta} \bar{Q}_{\dot\alpha i}~,
        \\
        & [K_{\alpha\dot{\alpha}}, \bar{Q}^i_{\dot{\beta}}] =- \varepsilon_{\dot\alpha\dot\beta}  S^{ i}_{\alpha}~,
        \quad [P_{\alpha\dot{\alpha}}, \bar{S}_{ \dot\beta i}] =- \varepsilon_{\dot\alpha\dot\beta} Q_{\alpha i}~,
   \end{split}
\end{equation}
\begin{equation}
\begin{split}
        &%[M_{mn}, P_s ] = i(\eta_{ns}P_m -\eta_{ms}P_n)~, \;
        [L_{(\alpha\beta)}, Q^i_\gamma]
        = -2 \varepsilon_{(\alpha\gamma}  Q^i_{\beta)}~,
        \quad
        [L_{(\alpha\beta)}, S^i_\gamma]
        = -2 \varepsilon_{(\alpha\gamma}  S^i_{\beta)}~,
   \end{split}
\end{equation}
\begin{equation}
\begin{split}
        &
        [T^i_j, Q^k] = \delta_j^k Q^i - \frac{1}{2} \delta^i_j Q^k,
        \quad
            [T^i_j, \bar{Q}_k] = -\delta_k^i \bar{Q}_j + \frac{1}{2} \delta^i_j \bar{Q}_k,
                \\
            &
            [T^i_j, S^k] = \delta_j^k S^i - \frac{1}{2} \delta^i_j S^k,
            \quad
            [T^i_j, \bar{S}_k] = -\delta_k^i \bar{S}_j + \frac{1}{2} \delta^i_j \bar{S}_k,
        \\
        & [T^{i}_{j}, T^{k}_{l}] = -i(\varepsilon^{ik} T_{jl} +  \varepsilon_{jl} T^{ik})~.
    \end{split}
\end{equation}
\end{subequations}
The generators of $SU(2)_{conf}$  satisfy the anti-Hermitian condition $(T^i_j)^\dagger = - T_i^j$.
The analyticity-preserving realization of the superconformal transformations in $\mathcal{N}=2$ HSS is well-known, see, e.g., \cite{18, Galperin:1985zv, Buchbinder:2024pjm, Buchbinder:2025ceg}.

%\medskip

The $OSp(2|4)$ spinorial generators can be composed from the spinorial $SU(2,2|2)$ generators as \cite{Gates:1983nr, Bandos:2002nn}\footnote{ Similar embeddings of superalgebras can be found in the literature. For example,  embedding of
superalgebra $su(2|1)$ into $D(2,1;\alpha)$ can be realized in a similar manner \cite{Ivanov:2013cea, Ivanov:2015iia}.}
\begin{equation}
\Psi^i_\alpha = Q^i_\alpha + c^{ik} S_{k\alpha}, \quad \bar{\Psi}_{\dot\alpha i} = \overline{\Psi^i_\alpha} = \bar{Q}_{\dot\alpha i} + c_{ik}\bar{S}^k_{\dot\alpha},
\end{equation}
where $c^{ik}$ are symmetric constants with dimension of mass, $[c^{ij}]=1$,
\bea
 c^{ik} = c^{ki}\,, \qquad \overline{c^{ik}} = c_{ik} = \varepsilon_{il}\varepsilon_{kj} c^{lj}. \nonumber
\eea
Using the algebraic relations \eqref{eq: N=2 SC a}, we obtain for the  ``holomorphic'' anticommutator:
\bea
\{ \Psi^i_\alpha,  \Psi^k_\beta \} = -4c^{ik} L_{(\alpha\beta)} +4 i \varepsilon_{\alpha\beta}\varepsilon^{ik} I, \qquad I := c_{lm}T^{lm}.
\label{holomorphic commutator}
\eea
Comparing this with the second line of \eqref{eq: N=2 SC a}, we observe  that $c^{ik}$ appear as the structure constants which completely break the scale symmetry,
$U(1)$ R symmetry and partially break $SU(2)_{conf}$ just to $SO(2) \sim U(1)$. Choosing the $SU(2)$ frame as
\be\label{eq: SU(2) frame}
c^{11} = c^{22},\, c^{12} =0\, \; \Leftrightarrow \; c^{ik} := \delta^{ik} m\,,
\ee
this anticommutator can be put in the more accustomed form,
\bea
\{ \Psi^i_\alpha,  \Psi^k_\beta \} = -4 m \delta^{ik} L_{(\alpha\beta)} +4 i \varepsilon_{\alpha\beta}\varepsilon^{ik} I\,.
\eea
Similarly, for the mixed anticommutator we get
\bea
&& \{ \Psi^i_\alpha,  \bar\Psi_{\dot\alpha k}\} = 4\delta^i_k \,P_{\alpha\dot\alpha} + 4 c^{ij}c_{k j} K_{\alpha\dot\alpha} := 4\delta^i_k\,R_{\alpha\dot\alpha}\,,
\eea
where in the right-hand side we encounter the generator of non-linear AdS translations:
\begin{equation}
    \begin{split}
&R_{\alpha\dot\alpha} = P_{\alpha\dot\alpha} + \frac12 c^2\, K_{\alpha\dot\alpha}\,, \quad c^2 := c^{ij}c_{ij} = 2m^2 \geq 0\,, \\
%\eea
%which satisfy the commutation relation:
%\bea
& [R_{\alpha\dot\alpha}, R_{\beta\dot\beta}] =
\frac12 c^2\big(\varepsilon_{\alpha\beta} \bar{L}_{(\dot\alpha\dot\beta)} + \varepsilon_{\dot\alpha \dot\beta} L_{(\alpha\beta)} \big).
\end{split}
\end{equation}

In the coordinate language, the above construction of spinor generators amounts to identifying the parameters of supersymmetry and conformal
supersymmetry\footnote{For what follows, it is useful to quote the conjugation rules of the spinor parameters,
$
    \bar\epsilon^i_{\dot\alpha} = \overline{\epsilon_{\alpha i}}\,,  \bar\eta_{\dot\alpha i} = \overline{\eta_{\alpha}^i}\,.$} as:
\begin{equation}\label{eq: id 1}
    \begin{split}
&\qquad\qquad\quad\; \eta^{\alpha i } \,\rightarrow c^{ik} \,\epsilon^\alpha_k         \,,
\quad \bar\eta^{\dot\alpha}_i \rightarrow c_{ik}\,\bar\epsilon^{\dot\alpha k}\,,  \\
& \epsilon^\alpha_i Q^{i}_\alpha + c^{ik}\epsilon_k^\alpha S_{\alpha i} = \epsilon^\alpha_i\Psi^i_\alpha\,, \quad
\bar \epsilon^i_{\dot\alpha} \bar{Q}_{i}^{\dot\alpha} + c_{il}\bar\epsilon_{\dot\alpha}^{l} \bar{S}^{\dot\alpha i} =  \bar \epsilon^i_{\dot\alpha}\bar{\Psi}^{\dot\alpha}_i\,,
\end{split}
\end{equation}
and the parameters of special conformal and  Poincar\'e translation transformations as:
\begin{equation}\label{eq: id 2}
    k_{\alpha\dot{\alpha}}  \rightarrow \frac{1}{2}c^2 a^{\alpha\dot{\alpha}},
    \qquad
    a^{\alpha\dot{\alpha}} P_{\alpha\dot{\alpha}}
    +
    k^{\alpha\dot\alpha} K_{\alpha\dot{\alpha}}  =     a^{\alpha\dot{\alpha}} R_{\alpha\dot{\alpha}}.
\end{equation}

To close this section, we present the total set of non-trivial (anti)commutation relations of superalgebra $\mathfrak{osp}(2|4)$:
\begin{equation}\label{eq: osp}
    \begin{split}
        &
        \{ \Psi^i_\alpha,  \bar\Psi_{\dot\alpha k}\} =  4\delta^i_k\,R_{\alpha\dot\alpha}\,,
        \\
        &\{ \Psi^i_\alpha,  \Psi^k_\beta \} = -4c^{ik} L_{(\alpha\beta)} +4 i \varepsilon_{\alpha\beta}\varepsilon^{ik} I,
        \\
        &
        [R_{\alpha\dot{\alpha}}, \Psi^i_\beta] = \varepsilon_{\alpha\beta} c^{ik} \bar{\Psi}_{\dot{\alpha}k},
        \quad
    [R_{\alpha\dot{\alpha}}, \bar{\Psi}^i_{\dot\beta}] =
    -\varepsilon_{\dot{\alpha}\dot{\beta}} c^{ik}\Psi_{\alpha k},
    \\
    & [R_{\alpha\dot\alpha}, R_{\beta\dot\beta}] = \frac12 c^2(\varepsilon_{\alpha\beta} \bar{L}_{(\dot\alpha\dot\beta)} + \varepsilon_{\dot\alpha \dot\beta} L_{(\alpha\beta)} ),
    \\
    & [I, \Psi^i] = - c^i_j \Psi^j,
    \quad
    [I, \bar{\Psi}^i] = c^i_j  \bar{\Psi}^j.
    \end{split}
\end{equation}
In the limit $c^{ij}\to0$ we reproduce $\mathcal{N}=2$ Poincar\'e superalgebra with the single central charge $Z = 2I$.

\medskip

\textbf{Remark on dS supersymmetry}

\smallskip

Since the $4D$ conformal algebra $\mathfrak{so}(2,4)\sim \mathfrak{su}(2,2)$ contains both the AdS algebra $\mathfrak{so}(2,3) \sim \mathfrak{sp}(4)$ and the dS
algebra $\mathfrak{so}(1,4)$ as its bosonic subgroups, one might expect that the dS superalgebra may be embedded in $\mathfrak{su}(2,2|2)$ like $\mathfrak{osp}(2|4)$.
The main obstacle to such a possibility  is that the $\mathfrak{so}(1,4)$ spinors are doubled compared to
 $\mathfrak{so}(1,3)$ spinors and so do not admit the Majorana condition. Though, an alternative symplectic Majorana condition can be imposed instead \cite{Lukierski:1984it, Pilch:1984aw}.
 It can be expected that, by imposing a similar (pseudo-conjugation) condition on composite generators $\Psi$, one could construct also the dS  superalgebra. However,  we deal here
 exclusively with the conventional conjugation, which distinguishes just the superalgebra $\mathfrak{osp}(2|4)$ as the only option.

\section{Hypermultiplet and realization of OSp(2|4) in HSS} \label{sec: 21}

The starting point of our study will be the standard free action of the massless hypermultiplet \cite{Galperin:1984av,18}:
\begin{equation}\label{eq:free hyper action}
    S_{free}= - \frac{1}{2} \int d\zeta^{(-4)}\, q^{+a} D^{++} q^+_a
    =
    - \int d \zeta^{(-4)}\; \tilde{q}^+ D^{++} q^+,
\end{equation}
where we used the notations:
\begin{equation}
D^{++} = \partial^{++} - 4i\theta^{+\alpha}\bar\theta^{+\dot\alpha}\partial_{\alpha\dot\alpha} \,, \quad   q^{+a} = (\tilde{q}^+, q^+) ,
    \qquad
    q^+_a = \epsilon_{ab} q^{+b} =\begin{pmatrix}
        q^+
        \\
        -\tilde{q}^+
    \end{pmatrix}.
\end{equation}
The superfield $q^{+a}$ is defined on the analytic harmonic superspace $\zeta := (x^{\alpha\dot\alpha}, \theta^{+\alpha}, \bar\theta^{+\dot\alpha}, u^\pm), \,q^{+a} = q^{+a}(\zeta)$.
It is a doublet with respect to the Pauli-G\"ursey group ${\rm SU}(2)_{PG}$.
The ${\rm SU}(2)_{PG}$ - covariant notation is  useful, when constructing the higher-spin vertices \cite{Buchbinder:2022kzl, Buchbinder:2022vra, Buchbinder:2024pjm}.

The above action is invariant under the whole ${\cal N}=2$ superconformal group and, hence, under its $OSp(2|4)$ subgroup. Explicitly,
the latter is realized on the analytic superspace coordinates as follows\footnote{These transformation are obtained from the $\mathcal{N}=2$ superconformal transformations in HSS
(see, e.g., Ref. \cite{Buchbinder:2024pjm}) through the identifications \eqref{eq: id 1} and \eqref{eq: id 2}.}

\smallskip

\textbf{1.} The \textbf{super AdS} transformation:
\begin{equation}
    \begin{split}
& \delta_\epsilon x^{\alpha\dot\alpha} = -4i\, \big[\epsilon^{\alpha i}\bar\theta^{+\dot\alpha} + \theta^{+\alpha}\bar\epsilon^{\dot\alpha i}
- c^{ik} \big(x^{\alpha\dot\beta}\bar\epsilon_{\dot\beta k} \bar\theta^{+\dot\alpha} + x^{\beta\dot\alpha}\epsilon_{\beta k}\theta^{+\alpha}\big)\big] u^-_i , \\
& \delta_\epsilon \theta^{+ \alpha}= \big(\epsilon^{\alpha i} - x^{\alpha\dot\alpha} c^{ik}\bar\epsilon_{\dot\alpha k}\big) u^+_i - 2i(\theta^+)^2 c^{ki}\epsilon^\alpha_k u^-_i\,,  \\
& \delta_\epsilon \bar\theta^{+ \dot\alpha}= \big(\bar\epsilon^{\dot\alpha i}+ x^{\alpha\dot\alpha} c^{ik}\epsilon_{\alpha k} \big) u^+_i
+ 2i(\bar\theta^+)^2 c^{ik}\bar\epsilon^{\dot\alpha}_k u^-_i,  \\
&
\delta_\epsilon u^{+i} = -4i \big[ c^{kl} u^+_k  (\epsilon_{\alpha l}\theta^{+ \alpha} + \bar\epsilon_{\dot\alpha l}\bar\theta^{+\dot\alpha})\big] u^{-i} \,.\lb{AdSsuper}
\end{split}
\end{equation}

\textbf{2.} The \textbf{nonlinear AdS translations}:
\begin{equation}\lb{NonlTran}
    \begin{split}
& \delta_{a} x^{\alpha\dot\alpha} = a^{\alpha\dot\alpha} + \frac12 c^2 a_{\beta\dot\beta}x^{\alpha\dot\beta} x^{\beta\dot\alpha}
= a^{\alpha\dot\alpha} \big(1 - \frac14 c^2 x^2\big) + \frac12 c^2(ax) x^{\alpha\dot\alpha}\,, \\
& \delta_{a} \theta^{+\alpha} = \frac12 c^2 a_{\beta\dot\beta} x^{\alpha\dot\beta} \theta^{+\beta}
=
\frac{1}{4} c^2 (ax) \theta^{+\alpha}
+
\frac{1}{2} c^2 x^{(\alpha \dot\beta} a_{\beta)\dot\beta} \theta^{+\beta} \,, \\
&
 \delta_{a} \bar\theta^{+\dot\alpha}
= \frac12 c^2 a_{\beta\dot\beta} x^{\beta\dot\alpha} \bar\theta^{+\dot\beta}
=
\frac{1}{4} c^2 (ax) \bar{\theta}^{+\dot{\alpha}}
+
\frac{1}{2} c^2 x^{\beta(\dot\alpha} a_{\beta\dot{\beta})} \bar{\theta}^{+\dot{\beta}}
\, , \\
& \delta_{a} u^{+i} = 2i  \left(c^2 a_{\alpha\dot\alpha} \theta^{+\alpha}\bar\theta^{+ \dot\alpha} \right) u^{-i}\,.
\end{split}
\end{equation}
These transformations can be obtained by computing the Lie bracket of two AdS supersymmetries, $(\delta_2\delta_1 - (1\leftrightarrow 2))$, with
$
a^{\alpha\dot\alpha} = 4i \big(\epsilon^{\alpha k}_{(1)}\bar\epsilon^{\dot\alpha}_{k (2)} - (1 \leftrightarrow 2)\big)
$.

\textbf{3.} The closure of AdS supersymmetry also contains \textbf{$SO(2)$ transformation} as a remnant of the conformal $SU(2)_{conf}$,
with parameter $\lambda^{(ij)} = \gamma c^{(ij)}$, $ \gamma = 4i\big(\epsilon^k_{\alpha (1)}\epsilon^\alpha_{k (2)} - c.c.\big)$:
\begin{equation}
    \begin{split}
& \delta_{\gamma} x^{\alpha\dot\alpha}  =
- 4i \gamma \, c^{--} \theta^{+\alpha} \bar\theta^{+\dot\alpha}\,, \\
& \delta_{\gamma} \theta^{+\alpha} = \gamma \,c^{+-}\theta^{+\alpha}\,,
\\
& \delta_\gamma \bar\theta^{+\dot\alpha}
= \overline{(\delta_\gamma \theta^{+\alpha})}
=
\gamma c^{+-} \bar{\theta}^{+\dot{\alpha}}\,, \\
&\delta_{\gamma} u^{+i} = \gamma\,  c^{++} u^{-i}, \lb{U1}
\end{split}
\end{equation}
where $c^{\pm\pm} := c^{ik}u^\pm_i u^\pm_k\,, \; c^{+-} = c^{-+} := c^{ik}u^+_i u^-_k\,$.

\smallskip

Note that, while checking $\mathfrak{osp}(2|4)$ invariance of various real expressions, it is enough to restrict attention only  to the holomorphic part of the AdS supersymmetry \eqref{AdSsuper}, with the parameters $\epsilon^k_\alpha$.
For further use, we quote such a holomorphic part of the transformation of the measure of integration over analytic superspace $d\zeta^{(-4)} =  d^4x d^4\theta du$:
\bea
\delta_{\epsilon} d\zeta^{(-4)} = 8i \left( c^{kl}u^-_k \epsilon_{\alpha l} \theta^{+\alpha}\right) d\zeta^{(-4)}. \lb{TranMes}
\eea
To compensate this variation in the action \eqref{eq:free hyper action} of $q^{+a}$, the latter should include the appropriate weight factor in its transformation
\be
\delta_{\epsilon} q^{+ a} = - 4i \left(c^{kl}u^-_k\epsilon_{\alpha l} \theta^{+\alpha} \right) q^{+a}. \lb{TranHyp}
\ee
Also, the holomorphic transformation of the flat harmonic derivative $D^{++} $ is given by
\be
\delta_\epsilon D^{++} =  -\lambda^{++}_\epsilon D^0 =  4i \left(c^{kl}u^+_k\epsilon_{\alpha l} \theta^{+\alpha}\right) D^0 \,, \lb{TranD++}
\ee
where $\delta_\epsilon u^+_i = \lambda^{++}_{\epsilon} u^-_i$, see eq. \eqref{AdSsuper}.

\section{Mass term for AdS hypermultiplet}\label{sec 3}

We are interested in the deformation of the action \eqref{eq:free hyper action} that preserves only AdS$_4$ supersymmetry but breaks the superconformal one.

One way to break $\mathcal{N}=2$ superconformal symmetry of massless hypermultiplet action \eqref{eq:free hyper action} is to introduce a massive term.
The standard way to do this is to introduce an auxiliary coordinate $x^5$ \cite{18, Ohta:1985ba} playing the role of central charge coordinate in $4D, \mathcal{N}=2$ Poincar\'e supersymmetry.
The hypermultiplet superfield is assumed to trivially depend on $x_5$:
\begin{equation}
    q^{+} (\zeta, x^5) = e^{-im_q x^5} q^{+}(\zeta).
\end{equation}
We then introduce cubic interaction of hypermultiplet with $\mathcal{N}=2$ vector supermultiplet:
\begin{equation}
    S_{int} = - \frac{1}{2} \int d\zeta^{(-4)}\, q^{+a} H^{++5} \partial_5 q^+_a =  im_q  \int d\zeta^{(-4)}\,  H^{++5} \tilde{q}^+ q^+.
\end{equation}
This interaction  is invariant under $\mathcal{N}=2$ superconformal symmetry.
Moreover, full hypermultiplet action $S_{free}+S_{int}$ is invariant under $x^5$ transformations $x^5 \to x^5 + \lambda^5(\zeta)$ accompanied by the abelian gauge transformations
\begin{equation}
    \delta_{\lambda^5} H^{++5} = D^{++} \lambda^5.
\end{equation}
Using this gauge freedom one can impose WZ-type gauge:
\begin{equation}\lb{WZ}
    H^{++5}_{WZ} = i (\theta^+)^2 \phi (x) - i (\bar{\theta}^+)^2 \bar{\phi}(x)
    - 4i \theta^{+\alpha} \bar{\theta}^{+\dot{\alpha}} A_{\alpha\dot{\alpha}}(x) + (\theta^+)^4 D^{ij}(x) u^-_i u^-_j
\end{equation}
with the residual gauge freedom $\delta_{res} A_{\alpha\dot{\alpha}} = \partial_{\alpha\dot{\alpha}} \lambda^5 \vert (x)$.

If we choose the $H^{++5}$ vacuum background as $\phi =  \bar\phi = 1$, all other fields in \eqref{WZ} possessing  zero vacuum values,
we regain just the central-charge extended ${\cal N}=2$ Poincar\'e supersymmetry as the invariant group of this vacuum, with the superconformal symmetry
being fully broken on such a vacuum and the quantity $m_q$ being a mass of hypermultiplet (see, e.g. Section 5.2.4 in \cite{18} and \cite{Buchbinder:2025ceg, Ohta:1985ba, Buchbinder:1997pw}). The operator  $\sim \partial_5$ is the central charge.

Now we wish to choose another background with broken $SU(2,2|2)$, such that it respects the properly realized AdS$_4$ supersymmetry only. This can be accomplished as follows.
We require that vacuum values of component fields
of $H_{WZ}^{++5}$ break superconformal symmetry to AdS supersymmetry\footnote{The superfield $H^{++5}$ can be interpreted as a vector compensator of $\mathcal{N}=2$ Einstein supergravity. Then vacuum values of
some fields from this supermultiplet compensate part of local superconformal transformations of $\mathcal{N}=2$ conformal supergravity. We postpone the detailed study of this scenario to the future publications.}
\eqref{AdSsuper}, \eqref{NonlTran} and \eqref{U1}, i.e.:
\begin{equation}
    \delta_{AdS} H^{++5}_{AdS} = D^{++} \lambda^5_{AdS}
\end{equation}
for some analytical parameter $\lambda^5_{AdS}(\zeta)$.
As the solution to this equation, we obtain the following expressions for $H^{++5}$,
\begin{equation}\label{eq: H^{++5} AdS}
    H^{++5}_{AdS} = i \left[(\theta^+)^2 -  (\bar{\theta}^+)^2\right] e(x)
    -
    6(\theta^+)^4 e(x)^2 \,c^{(ij)}u^-_i u^-_j,
\end{equation}
and for $\lambda^5$ (up to the harmonic-independent part):
\begin{equation}\label{eq: lambda5 susy}
    \begin{split}
        \lambda^5_a = \,&0,
        \\
        \lambda^5_{\epsilon} = \,&
       % 2i e(x) \Big[ (\theta^+ \epsilon^-)
       % -
       % x^{\alpha\dot{\alpha}} \eta^-_\alpha \bar{\theta}^+_{\dot{\alpha}}  \Big]
       % e(x) \Big\{1 - 2i \left[(\theta^+)^2 - (\bar{\theta}^+)^2 \right] e(x)  c^{--}  \Big\} + (c.c.),
        2i e(x) \Big[ (\theta^+ \epsilon^-) - x^{\alpha\dot{\alpha}}\bar{\theta}^+_{\dot{\alpha}}( \epsilon^+_\alpha c^{--} - \epsilon^-_\alpha y)\Big]
     \Big\{1 - 2i \left[(\theta^+)^2 - (\bar{\theta}^+)^2 \right] e(x)  c^{--}  \Big\} + (c.c.),
        \\
        \lambda^5_{\gamma} = \,& i  \gamma c^{--} e(x)\Big( \left[(\theta^+)^2 -  (\bar{\theta}^+)^2\right]
        -
        6  (\theta^+)^4 c^{--} e(x)\Big).
    \end{split}
\end{equation}
Here we used the short-cut notations $y = c^{ij} u^+_iu^-_j$, $e(x) := \frac{1}{1+ m^2x^2/2}$. In the limit $c^{(ij)}\to 0$
we restore the rigid $\mathcal{N}=2$ Poincar\'e supersymmetry mass term mentioned above:
\begin{equation} H^{++5}_{flat} =  i \left[(\theta^+)^2 -  (\bar{\theta}^+)^2\right],
    \qquad \lambda^5_\epsilon|_{flat} = 2i \left[ (\theta^+ \epsilon^{-})  - (\bar{\theta}^+ \bar{\epsilon}^{-})  \right].
\end{equation}

\medskip

As a result, $\mathcal{N}=2$ AdS transformations in the presence of the auxiliary coordinate $x^5$ can be represented by the analyticity-preserving differential operator:
\begin{equation}
    \hat{\Lambda}_{AdS} = \lambda_{AdS}^{\alpha\dot{\alpha}}\partial_{\alpha\dot{\alpha}} + \lambda^{\hat{\alpha}+}_{AdS} \partial^-_{\hat{\alpha}}
    +
    \lambda^{++}_{AdS} \partial^{--}
    +
    \lambda^5_{AdS} \partial_5,
\end{equation}
where the analytic parameters can be read off from eqs. \eqref{AdSsuper}, \eqref{NonlTran}, \eqref{U1} and \eqref{eq: lambda5 susy}.
The operator $\hat{\Lambda}_{AdS}$ is Killing vector field, which generate $OSp(2|4)$ isometries of $\mathcal{N}=2$ analytic AdS superspace.

\smallskip

Lie bracket of the coordinate transformations $(\delta_2 \delta_1 - (1\leftrightarrow2))$  will be modified only by the ``central charge'' terms, {\it viz.}
\begin{equation}
    \left(\hat{\Lambda}_{AdS}(2) \lambda^5_{AdS}(1) - (1\leftrightarrow2) \right) \partial_5,
\end{equation}
which leads to the  final formula for $\lambda^5_\gamma$:
\begin{equation}
 \lambda^5_\gamma =
 \frac{\gamma}{2}
 +
  i \left[(\theta^+)^2 -  (\bar{\theta}^+)^2\right]   \gamma c^{--} e(x)
 -
 6  (\theta^+)^4\gamma   \left( c^{--} \right)^2  e(x)^2.
\end{equation}
We see that in the flat limit the $SO(2)$ generator is reduced to the central charge generator,  $I \to \frac{1}{2} Z$,
\begin{equation}
    \{ \Psi^i_\alpha,  \Psi^k_\beta \} = -4c^{ik} L_{(\alpha\beta)} +4 i \varepsilon_{\alpha\beta}\varepsilon^{ik} I
    \qquad\quad
    \xrightarrow[c^{ik}\to 0]
    \qquad\qquad
        \{ \Psi^i_\alpha,  \Psi^k_\beta \} = 2 i \varepsilon_{\alpha\beta}\varepsilon^{ik} Z.
\end{equation}
This highlights (at least for $\mathcal{N}=2$ case) the relationship between the phenomenon of multiplet shortening for  extended Poincar\'e supersymmetry with central charges
and that for $OSp(\mathcal{N}|4)$ supersymmetry \cite{Freedman:1983na}.

\smallskip

A similar idea has already been applied to the construction of  massive hypermultiplet theories on the $AdS^{4|8}$ background within the framework of projective superspace \cite{Kuzenko:2008qw}. However in contrast to this
approach, in  our approach chiral field strength (which has the homogenous transformation law under $OSp(2|4)$ supersymmetry) has the component content\footnote{Full component content of $\mathcal{W}$ can be obtained using
solutions of zero curvature equation, which is presented e.g. in \cite{Ivanov:2024gjo}.  }
\begin{equation}
    \mathcal{W} = (\bar{\mathcal{D}}^+)^2 H^{--5}_{AdS} \sim ie(x) + (\theta^+)^2 e(x)^2 c^{ij} u_i^-u_j^- + \dots
\end{equation}
which does not satisfy the condition $\mathcal{W}=1$, which were used in Ref. \cite{Kuzenko:2008qw}.  This indicates that the comparison of the two approaches is not so straightforward.

\medskip

The conclusion is that it is possible to gain an ``external'' mass term for $q^{+}_{a}$ by extending
the coordinate action of generators of AdS supersymmetry  and  of the ``internal'' $SO(2)$ symmetry
through adding to them extra ``matrix'' pieces of an external $SO(2)$ realized on the hypermultiplet superfield.
 In the flat limit such modified ${\cal N}=2$ AdS  supergroup  contracts just into the centrally-extended
$4D, {\cal N}=2$ supersymmetry.  In this limit,  $SO(2)$ generator coincides with  $U(1)$ generator from $SU(2)_{PG}$ acting on the doublet indices of $q^{+} _{a}$.
To avoid a possible misunderstanding, we point out that the dependence on the extra coordinate $x_5$ was
introduced above just for convenience and in order to point out striking analogies with the central charges
in ${\cal N}=2$ Poincar\'e supersymmmetry. In fact, we could from the very beginning
identify $\partial_5$ with the matrix generator of the proper $SO(2) \subset SU(2)_{PG}$.

%As result we obtain central-charge extension of "holomorphic" anticommutator of $\mathfrak{osp}(2|4)$ superalgebra \eqref{eq: osp}:
%\begin{equation}
%   \{ \Psi^i_\alpha,  \Psi^k_\beta \} = -4c^{ik} L_{(\alpha\beta)} +4 i \varepsilon_{\alpha\beta}\varepsilon^{ik} \Big( I + \frac{1}{2} Z\Big).
%\end{equation}
%Generator $Z$ commute with all  $\mathfrak{osp}(2|4)$ generators.
%Other $\mathfrak{osp}(2|4)$ relations while not being modified.

%\newpage
\begin{flushleft}
    \textbf{The component contents}
\end{flushleft}
In order to study the component structure of the massive AdS hypermultiplet action we need to eliminate an infinite number of the auxiliary fields
present in the analytic superfield $q^{+a}$.
%irrespective of on which background $q^{+a}$ is considered.
The easiest way is to make use of the hypermultiplet equations of motions:
\begin{equation}
    D^{++} q^+ - i m_q H^{++5}_{AdS} q^+ = 0.
\end{equation}
After elimination of the auxiliary fields we obtain in the \textit{bosonic sector}:
\begin{equation}\label{eq: q on-shell}
    \begin{split}
        q_{on-shell}^+ =& f^iu^+_i - m_q \left[ (\theta^+)^2  - (\bar{\theta}^+)^2  \right] e(x) f^i u^-_i
        \\&+
        4i \theta^{+\alpha}\bar{\theta}^{+\dot{\alpha}} \partial_{\alpha\dot{\alpha}} f^i u^-_i
        +
        2im_q (\theta^+)^4 e(x)^{2} c^{(ij} f^{k)}\, u^-_{(i} u^-_j u^-_{k)}.
    \end{split}
\end{equation}
Then the resulting on-shell $q^{+a}$ action in the bosonic sector is reduced to:
\begin{equation}
    \begin{split}
        S_{scalar} &=
        \int d^4x \; \Big(  \partial_n f^i \partial^n \bar{f}_i  - m_q^2 e(x)^2 f^i \bar{f}_i
        -
     im_q e(x)^2 f^i \bar{f}^j c_{(ij)}
        \Big).
    \end{split}
\end{equation}

Under non-linear AdS translations \eqref{NonlTran} the doublet of scalar fields $f^i$ transforms as:
\begin{equation}
    \delta_a f^i(x) = - m^2 (a x) f^i(x).
\end{equation}
After redefinition $f^i \to e(x) \hat{f}^i$ we obtain AdS covariant scalar field, $\delta_a \hat{f}^i (x) = 0$.
Then the first two terms give the standard kinetic action for the massive scalars given on AdS metric $g_{mn} =
e(x)^2 \eta_{mn}$ (with the scalar curvature  $R = 4\Lambda = -96m^2 = -48 c^{ij}c_{ij}$):
\begin{equation}
    S_{scalar} = \int d^4x \sqrt{-g} \left( g^{mn} \partial_m \hat{f}^i \partial_n\bar{\hat{f}}_i
    +
    \frac{R}{6} \hat{f}^i \bar{\hat{f}}_i
    - m^2_q \hat{f}^i \bar{\hat{f}}_i
        -
    im_q  \hat{f}^i \bar{\hat{f}}^j c_{(ij)} \right).
\end{equation}
 The last term provides the mass splitting for the complex scalar doublet. To make this point more visual it is convenient to pass to the special $SU(2)$ frame $c_{ij} = m \delta_{ij}$ \eqref{eq: SU(2) frame}. In this frame,
 after  diagonalization of the mass matrix $m^2_q \epsilon_{ij} + i m_q c_{(ij)}$, we recover masses of the scalar hypermultiplet fields as:
\begin{equation}\label{eq: bos masses}
    m_\pm^2 = m_q^2 \pm m m_q.
\end{equation}
Note that the scalar masses automatically satisfy the well-known Breitenlohner-Freedman bound \cite{Breitenlohner:1982bm},
$m_\pm^2 = (m_q \pm \frac{m}{2})^2 - \frac{m^2}{4} \geq - \frac{m^2}{4}$ .

Analogously, the fermionic sector of the on-shell hypermultiplet looks as,
\begin{equation}
    q^+_{on-shell} = \theta^{+\alpha} \psi_\alpha
    +
    \bar{\theta}^+_{\dot{\alpha}} \bar{\kappa}^{\dot\alpha},
\end{equation}
and we obtain the on-shell fermionic action in the form:
\begin{equation}
    S_{fer} = \int d^4x\, \left( i \bar{\psi}^{\dot{\alpha}} \partial_{\dot\alpha}^\alpha \psi_\alpha +
    i \bar{\kappa}^{\dot{\alpha}} \partial_{\dot\alpha}^\alpha \kappa_\alpha
    +
    e(x)
    \frac{m_q}{2} \psi^\alpha \kappa_\alpha
    +
    e(x)
    \frac{m_q}{2} \bar{\psi}_{\dot\alpha} \bar{\kappa}^{\dot\alpha}
    \right).
\end{equation}

Under non-linear AdS translations fermionic fields transform according to
\begin{equation}
    \begin{split}
    &\delta_a \psi_\alpha(x) = - \frac{3}{2} m^2 (ax) \psi_\alpha(x) - \ell_{(\alpha}^{\;\;\beta)} \psi_\beta(x),
    \\
    &
    \delta_a \bar{\psi}_{\dot{\alpha}}(x)
    =
    - \frac{3}{2} m^2 (ax)  \bar{\psi}_{\dot{\alpha}}(x) - \bar{\ell}_{(\dot{\alpha}}^{\;\;\dot{\beta})}  \bar{\psi}_{\dot{\beta}}(x),
    \end{split}
\end{equation}
where the induced Lorentz parameters $\ell_{(\alpha\beta)}$ and $\bar{\ell}_{(\dot{\alpha}\dot{\beta})}$ are explicitly given  in \eqref{eq: nonl Lorentz}.
 Similarly to scalar fields, we can redefine $\psi_\alpha(x) \to e(x)^{\frac{3}{2}} \hat{\psi}_\alpha(x)$ and, using the AdS covariant derivative \eqref{AdScovx}, obtain the action for Dirac spinor on AdS space:
\begin{equation}
    S_{fer} = \int d^4x \sqrt{-g} \left\{
     i \bar{\hat{\psi}}^{\dot{\alpha}} \left(\nabla_{\dot\alpha}^\alpha - \frac{3}{2} m^2 x^\alpha_{\dot{\alpha}} \right) \hat{\psi}_\alpha +
    i \bar{\hat{\kappa}}^{\dot{\alpha}} \left(\nabla_{\dot\alpha}^\alpha - \frac{3}{2} m^2 x^\alpha_{\dot{\alpha}} \right)  \hat{\kappa}_\alpha
    + \frac{m_q}{2}\big(\hat{\psi}^\alpha \hat{\kappa}_\alpha
    + \bar{\hat{\psi}}_{\dot\alpha} \bar{\hat{\kappa}}^{\dot\alpha}\big) \right\}.
\end{equation}
So we conclude,  that the  fermions have masses  $m_f = m_q$, which differ from boson  masses \eqref{eq: bos masses}. This is a general property of AdS supermultiplets (see, e.g.,
\cite{Ivanov:1980vb, IvSor2} for the  ${\cal N}=1$ case). The masses satisfy the simple sum rule
\begin{equation}
    m_+^2 + m_-^2 = 2m^2_f.
\end{equation}

\section{Harmonic Weyl rescaling}\label{sec 4}

One of the most striking and effective achievements of the harmonic superspace method  was the construction
of general hyper-K\"ahler (HK) sigma  models on flat superspace \cite{18, Galperin:1985de}. It is
interesting to address this problem in the case of $\mathcal{N}=2$ AdS supersymmetry. The first step towards this goal
is to define an analytic super AdS integration measure, which would transform so as to compensate the weight factor \eqref{TranMes}
in the transformation law  of the standard measure (see below).

The analytic superspace integration measure is transformed under the AdS supersymmetry as (we are interested in the holomorphic part of the whole transformation):
\bea
\delta_\epsilon d\zeta^{(-4)} = 8i \,(c^{kl}u^-_k \epsilon_{\alpha l} \theta^{+\alpha}) d\zeta^{(-4)} = -8i \,(c^{+-} \epsilon^{-\alpha} -c^{--}\epsilon^{+\alpha})\theta^+_\alpha\,d\zeta^{(-4)}. \lb{TranMes}
\eea
We are going to define the scalar factor with the transformation law compensating
this non-invariance\footnote{Note that  in Ref. \cite{Butter:2015nza}
an  analytic density in the conformal $\mathcal{N}= 2$ supergravity was introduced such that it compensates the whole gauge ${\cal N}=2$ superconformal  transformation
of the flat analytic integration measure.  Its possible relation to our AdS supergroup invariant measure is not clear for us. The measure $(Q^{+i}u^-_i)^2$ (here $Q^{+i}$
is the hypermultiplet compensator)  in $\mathcal{N}=2$ supergravity which was introduced in \cite{Galperin:1987ek}  (see also \cite{Bagger:1987rc})  is not invariant  under $x^5$ gauge transformations,
which is necessary for description of $\mathcal{N}=2$ supergravity
with cosmological constant, see discussion in \cite{18, Ivanov:2022vwc}.
So it also cannot be the prototype of our AdS measure, which contains no  $x^5$ dependence.}
%It is worth noting that the analytic measure $(Q^{+i}u^-_i)^2$ (here $Q^{+i}$ is the hypermultiplet compensator) in $\mathcal{N}=2$ supergravity
%introduced in \cite{Galperin:1987ek} (see also \cite{Bagger:1987rc}) is not invariant under $x^5$ gauge transformations, which is necessary for description of $\mathcal{N}=2$ supergravity
%with cosmological constant, see discussion in \cite{18, Ivanov:2022vwc}. In this connection,  it is important to emphasize that  in Ref. \cite{Butter:2015nza}
%the analytic density in the conformal $\mathcal{N}= 2$ supergravity was introduced. However, it is unclear how to use it for practical applications. }
\begin{equation}
\Sigma = \exp \Omega, \qquad \delta_\epsilon \Omega =  8i \, (c^{+-} \epsilon^{-\alpha} -c^{--}\epsilon^{+\alpha})\theta^+_\alpha\,. \label{Compens}
\end{equation}
To find $\Omega$, we parametrize it as
\bea
\Omega = F + c^{--}\big[(\theta^+)^2 l + (\bar \theta^+)^2 \bar l \big] + i c^{--} x_{\alpha\dot\alpha} \theta^{+\alpha}\bar\theta^{+\dot\alpha} K + (c^{--})^2 (\theta^+)^2(\bar\theta^+)^2 S\,,\lb{Omega}
\eea
where all the coefficients are as yet arbitrary functions of the variables $z:= x^{\alpha\dot{\alpha}}x_{\alpha\dot{\alpha}}$ and $y:= c^{+-}$  and satisfy reality conditions:
\begin{equation}\lb{RestronFKS}
	\begin{split}
		&F = F(z, y)\,, \quad l = l(z, y)\,, \quad \bar l = \bar l (z, y), \quad K = K(z, y)\,, \quad S = S(z, y)\,,
		\\
		&\bar F = F, \quad  \bar K = K, \quad \bar S = S.
	\end{split}
\end{equation}
Then we consider the supersymmetric variations of these coefficients under \eqref{AdSsuper}, which yields the
 equations\footnote{Some of these equations originally appear with the factors $c^{--}$ and $(c^{--})^2$. Then, for deriving the equations below, we need to multiply
 	their original forms by $c^{++}$ and $(c^{++})^2$. The new factors are non-singular, in the sense that their harmonic expansions start with constants,
 	$\int du c^{++}c^{--} \neq 0\,, \int du (c^{++}c^{--})^2 \neq 0\,.$ So one can divide by them to gain the final form of the equations.}  for the coefficients $F, l, K, S$:
 \begin{equation}
 	\begin{split}
 		c^{--}(\theta^+ \epsilon^+):\qquad & (a)\;\;\big(z\partial_z - y\partial_y\big) F -\frac{i}{2} l - \frac{1}{8} y z K = -2\,,  \\
 		(\theta^+ \epsilon^-):\qquad& (b)\;\;y z\partial_z F - (y^2 + m^2) \partial_y F - \frac1{8} z (y^2 + m^2) K = -2 y\,, \\
 		x_{\alpha\dot{\alpha}}\epsilon^{-\alpha} \bar{\theta}^{+\dot{\alpha}}:\qquad &
 		(c)\;\;\partial_z F + i \frac{l}{4} (y^2 + m^2) = 0\,, \quad \partial_z F - i \frac{\bar l}{4} (y^2 + m^2) = 0,\\
 		c^{--}x_{\alpha\dot{\alpha}}\epsilon^{+\alpha} \bar{\theta}^{+\dot{\alpha}}:\qquad& (d) \;\; 2y l - iK =0\,, \quad 2y \bar l + iK =0\,, \lb{4first}
 	\end{split}
 \end{equation}
 and
 \begin{equation}
 	\begin{split}
 		(c^{--})^2 (\bar{\theta}^+)^2 (\theta^+ \epsilon^+):
 		\quad
 		& (a)\;\;2S - 4i(z\partial_z - y\partial_y) l = 2S + 4i(z\partial_z - y\partial_y) \bar l = 0\,, \\
 		c^{--} (\bar{\theta}^+)^2 (\theta^+\epsilon^-):\quad
 		& (b)\;\; K + \frac12 z\partial_z K - iy z\partial_z \bar l + i (y^2 + m^2) \partial_y \bar l = 0
 		%\quad (\text{and c.c.})
 		\,,
 		\\
 		(c^{--})^2 (\theta^+)^2 x_{\alpha\dot{\alpha}}\epsilon^{+\alpha} \bar{\theta}^{+\dot{\alpha}}:\quad
 		& (c) \;\;K + (z\partial_z - y\partial_y)K + y S = 0\,,
 		\\
 		c^{--} (\theta^+)^2 x_{\alpha\dot{\alpha}} \epsilon^{-\alpha} \bar{\theta}^{+\dot{\alpha}}:\quad  & (d)\;\;(y^2 + m^2) S  + y K + y z\partial_z K - (y^2 + m^2) \partial_y K+ 4i\partial_z l = 0.
 		% \quad (\text{and c.c.})\,,
 		\lb{4last}
 	\end{split}
 \end{equation}
 The set \eqref{4first} is basic, while eqs. \eqref{4last} serve to express the function $S$ and provide
 some self-consistency checks for solutions of \eqref{4first}.

It is easy to find that
\begin{subequations}
\begin{equation}
F = 2 \log  e(x) + \log \left(1 + \frac{y^2}{m^2}\right), \lb{F2}
\end{equation}
\begin{equation}
l = i L\,, \quad K = 2 y L\,, \quad S = -2(z\partial_z - y\partial_y)L,
\quad
L = -4 m^2 e(x)\frac{1}{y^2 + m^2}\,. \lb{L}
\end{equation}
\end{subequations}

Then the analytic measure reads
\begin{equation}
    \begin{split}
    \Sigma =& \left(1 + \frac{y^2}{m^2}\right) e(x)^2%\\
%   &
-
    4i c^{--}  \big[(\theta^+)^2  - (\bar \theta^+)^2 + 2 y x_{\alpha\dot\alpha} \theta^{+\alpha}\bar\theta^{+\dot\alpha}\big] e(x)^3
    \\&
    +
    16 (c^{--})^2 (\theta^+)^4 e(x)^4 \left( 1- \frac{m^2}{4} z \right).
    \end{split}
\end{equation}

One of the unexpected properties of the analytic measure $\Sigma$ is lacking of the naive flat limit:
\begin{equation}
    \lim_{c^{ij}\to 0}\Sigma =  1 + \frac{c^{(ij)} c^{(kl)}}{c^2} u^+_i u^-_j u^+_k u^-_l
    =
    \frac{5}{6}
    +
    \frac{c^{(ij)} c^{(kl)}}{c^2} u^+_{(i} u^+_j u^-_k u^-_{l)}.
\end{equation}
In other words, there survives a contribution from the pure angular part of the constant triplet $c^{ik}$, {\it i.e.}, $\hat{c}^{ik} := c^{ik}/|c|$.
In fact, in order to correctly take the flat limit of the superfield Lagrangians it seems necessary to do before the integrals over harmonic variables.
We postpone discussion of this subtlety to a more detailed article.

\medskip

Let us now redefine the hypermultiplet $q^{+}_{a}$ as
\bea
q^{+}_{a} = \Sigma^{\frac12} \hat{q}^{+}_{a},
\qquad
\delta_{AdS} \hat{q}^{+}_{a} = 0\,.
\eea
The action \label{eq:hyper action} with manifest ${\cal N}=2$ Poincar\'e supersymmetry is transformed into the action
\bea
S_{free}^{AdS}= - \frac{1}{2} \int d\zeta^{(-4)}\,\Sigma  \; \hat{q}^{+a} D^{++} \hat{q}^+_a\,, \label{AdShyperaction}
\eea
which is \textit{manifestly invariant under ${\cal N}=2$ AdS supersymmetry}. Like its flat analog, it still possesses a hidden superconformal symmetry,
the precise realization of which in \eqref{AdShyperaction} is of no interest for our presentation here.

The existence of equivalent representations of the same superconformally invariant action of massless $q^{+a}$ in the analytic HSS with either manifest
${\cal N}=2$ Poincar\'e supersymmetry or manifest ${\cal N}=2$ AdS supersymmetry related by a super Weyl scalar factor means just
the generalized superconformal flatness of this analytic superspace. This property is quite analogously to the superconformal flatness of ordinary AdS superspaces \cite{Bandos:2002nn}.

Using $\hat{q}^+_a$ one can construct interaction Lagrangian ${\cal L}^{+4}(\hat{q}^{+ a}, u^\pm)$
which breaks conformal (and Poincar\'e ) supersymmetry of free action \eqref{AdShyperaction},  but still preserves ${\cal N}=2$ AdS supersymmetry. It is expected to yield, in its bosonic sector,
a kind of deformed hyper K\"ahler sigma model on $AdS_4$ background. Some further relevant details are sketched in section \ref{eq: sec concl}.

\medskip

For the component analysis (to be performed in a more detailed paper) it is useful to deal with the manifestly AdS covariant component fields. In Appendix \ref{app B}
we present the possible redefinitions of Grassmann coordinates leading to AdS covariant fields in the component expansions.

\section{$\mathcal{N}=2$ AdS massive vector multiplet}

As an interesting consequence of existence of the analytic AdS integration measure, one can construct the harmonic action for massive $\mathcal{N}=2$ AdS vector multiplet:
\begin{equation}
    S = \int d^4x d^8\theta du \, V^{++} V^{--} -m_V^2 \int d\zeta^{(-4)} \Sigma\,  V^{++} V^{++}.
\end{equation}
Here $V^{--}$ is defined as the solution of  zero-curvature equation $\mathcal{D}^{++} V^{--} = \mathcal{D}^{--} V^{++}$ with flat harmonic derivatives $\mathcal{D}^{\pm\pm} := \partial^{\pm\pm} - 4i \theta^{\pm\alpha} \bar{\theta}^{\pm\dot{\alpha}} \partial_{\alpha\dot{\alpha}} + \theta^{\pm\alpha} \partial^\pm_\alpha
    +
    \bar{\theta}^{\pm\dot{\alpha}} \bar{\partial}^\pm_{\dot{\alpha}}$.
The first (kinetic) term, as well as the flatness condition, are invariant under $\mathcal{N}=2$ superconformal symmetry.
The mass term breaks this symmetry to $\mathcal{N}=2$ AdS$_4$ supersymmetry, which is realized as $\delta_{AdS} V^{++} = 0$
and  $\delta_{AdS} V^{--} = - (\mathcal{D}^{--}\lambda_{AdS}^{++}) V^{--}$ (where $\delta_{AdS} u^+_i = \lambda_{AdS}^{++} u^-_i$).

\section{Conclusions}\label{eq: sec concl}

In this paper, we  for the first time studied the realization of $\mathcal{N}=2$ AdS supergroup $OSp(2|4)$ in harmonic superspace,
starting from the known analyticity-preserving realization of ${\cal N}=2$ superconformal group $SU(2,2|2)$. Thus, our construction ensures automatic
preservation of Grassmann ${\cal N}=2$ analyticity in the super AdS case like in the super Minkowski case.
We have constructed the AdS hypermultiplet mass term and the invariant integration measure in the analytic harmonic AdS$_4$  superspace.
These findings open a few directions for the future study.

\smallskip

$\bullet$ \textit{$\mathcal{N}=2$  supergravity  in AdS HSS background}

\smallskip

As is well known, $\mathcal{N}$-extended AdS superspace is superconformally flat \cite{Bandos:2002nn}
(see also \cite{Koning:2024iiq} for a recent discussion). The principal version of $\mathcal{N}=2$ supergravity can be formulated
in HSS in terms of $\mathcal{N}=2$ Weyl supermultiplet and two compensating multiplets, the vector
 one and \linebreak hypermultiplet \cite{18, Galperin:1987ek, Ivanov:2022vwc}.
It is instructive to explicitly find all vacuum values of these superfields defining $\mathcal{N}=2$ AdS background and to
construct the complete set of AdS covariant superderivatives, proceeding exclusively from the analytic harmonic superspace geometry
as a generalization of what has been done here to the case of local  $\mathcal{N}=2$ AdS supergroup.

\smallskip

$\bullet$ \textit{Coset approach to $OSp(2|4)$ supersymmetry}

\smallskip

The full-fledged superfield formulation of  $OSp(1|4)$ supersymmetry was developed in  ref.  \cite{Ivanov:1980vb}, where the ${\cal N}=1$ AdS superspace
was identified with the supercoset $OSp(1|4)/O(1,3)$. It seems important to consistently work out  an analogous formulation for $OSp(2|4)$ supersymmetry as well.
Analogously to the realization of $\mathcal{N}=2$ superconformal group in harmonic superspace \cite{18} (see also \cite{Galperin:1992pj, Butter:2015nza}),
in such a construction the harmonic variables must be associated with some extra $SU(2)_A$, and so one is led to consider the coset superspace
\begin{equation*}
\mathbb{HADS}^{4|8} =   \frac{OSp(2|4)}{O(1,3)\times SO(2)} \times \frac{SU(2)_A}{U(1)}
\end{equation*}
as the appropriate framework. We expect that this line of research would allow to get a deeper insight into the geometric structure
of the analytic harmonic AdS$_4$ superspace,
in both rigid and local cases.

%\smallskip

%$\bullet$ Component content? The meaning of the absence of a flat limit for a measure??

\smallskip

$\bullet$ \textit{AdS hyper K\"ahler sigma   models }
\smallskip

Of the considerable interest are also $\mathcal{N}=2$ AdS nonlinear sigma models, which seem to have drastic differences from those in $\mathcal{N}=2$ Minkowski case.
In particular, there are formulations where $\mathcal{N}=2$ supersymmetry algebra closes off shell without need in an infinite number of auxiliary fields.
Such models were constructed  in the projective $\mathcal{N}=2$ AdS superspace by  Kuzenko and Tartaglino-Mazzucchelli \cite{Kuzenko:2008qw}
and  then in $\mathcal{N}=1$ AdS superspace by  Butter and Kuzenko \cite{Butter:2011zt}.  These results were further developed
in Refs. \cite{Butter:2011kf, Butter:2012jj}.

Our approach allows to construct sigma models which are manifestly invariant under
$\mathcal{N}=2$  AdS supersymmetry. The corresponding analytic superfield Lagrangian for $n$ hypermultiplets $\hat{q}^+_a$ ($a =1,...,2n$)  has the form:
\begin{equation}
    S^{AdS}_{HK} = \int d\zeta^{(-4)} \Sigma \, \left[ \hat{q}^+_a D^{++} \hat{q}^{+a} + L^{(+4)} (\hat{q}^+, w^+, u^-) \right],\lb{AdSHK}
\end{equation}
where
\be
w^{+i} := u^{+ i} - u^{- i} c^{++} \frac{y}{y^2 + m^2}\,, \qquad \delta_{AdS}w^{+i} = 0\,, \lb{Newharm}
\ee
is a new harmonic variable inert under $OSp(2|4)$ (the harmonic derivative $D^{++}$ still contains differentiation with respect to the
ordinary harmonics $u^{\pm i}$ only and $\hat{q}^{+a}$ also depends on $u^{\pm i}$). The action \eqref{AdSHK} is the AdS  generalization of the general $q^{+}$ action in the Minkowski background
which  yields the most general hyper-K\"ahler sigma model in the bosonic sector. It is interesting that at this point we once again encounter
the problem of the correct implementation of the flat Minkowski space limit. The coefficient of $u^{-i}$ in the
second piece in \eqref{Newharm} can be rewritten as
\be
c^{++}\frac{y}{y^2 + m^2} = \hat{c}^{++}\,\frac{y}{|m|}\,\left(1 + \frac{y^2}{m^2}\right)^{-1}\,, \quad \hat{c}^{ik} = \frac{c^{ik}}{|m|}\,,
\ee
so that only the dimensionless angular part of $c^{ik}$ appears in \eqref{Newharm}.

One could avoid this subtlety by retaining, in the potential in \eqref{Newharm}, the explicit dependence only on $u^{-i}$, thereby narrowing the admissible
class of HK sigma models. It seems, however, that the unremovable presence of the angular part of the $SU(2)$ breaking parameter $c^{ik}$ in the general
${\cal N}=2$ sigma model action of the AdS$_4$ hypermultiplet may be responsible for an essential difference in the geometry of ${\cal N}=2$ sigma models in Minkowski
and AdS superspaces. It is worth noting in this connection that the non-analyticity of passing from the AdS sigma model actions to their Minkowski counterparts
was also pointed out in \cite{Butter:2011zt}.
However, the action \eqref{AdSHK} certainly involves an infinite number of auxiliary fields through the unconstrained analytic $\hat{q}^{+a}$ and for this reason the
general target geometry of the bosonic sector of \eqref{AdSHK} should be radically different from the one discussed in \cite{Butter:2011zt}. The
superfield potentials in \eqref{AdSHK}  generically do not exhibit any external isometry, quite similar to its flat ${\cal N}=2$
supersymmetry counterparts \cite{18} (there should be present of course the internal $SO(2)$ symmetry as the necessary part of the underlying $OSp(2|4)$ supersymmetry).
Anyway, it would be interesting to understand in full the interplay between the constant triplet $c^{ik}$ specifying the AdS world-volume geometry and the relevant target sigma model geometries.
The HSS approach seems to provide the most natural framework for such an analysis in view, e.g.,
of the fact that the projective ${\cal N}=2$ superspace is a particular case of HSS \cite{Sergei}.

We hope to explore the properties of such models in more detail and to make comparison between different approaches elsewhere.

\smallskip

$\bullet$ \textit{$\mathcal{N}=2$ AdS supergravity: analytic prepotentials and linearized action}

\smallskip

$\mathcal{N}=2$ AdS supergravity was previously elaborated in \cite{Butter:2011ym} based on papers \cite{Butter:2010jm, Butter:2010sc}
in terms of both ordinary $\mathcal{N}=2$ superfields and $\mathcal{N}=1$ superfields.
However, the formulation in harmonic superspace, the role of Grassmann analyticity, the structure of prepotentials and gauge group have never been fully elucidated.
The structure of analytic prepotentials and the gauge group of  $\mathcal{N}=2$
supergravity in HSS have a beautiful geometric interpretation (see \cite{Ivanov:2022vwc} for a recent review) and allow a natural generalization to $\mathcal{N}=2$ higher-spin theories.
We expect that similar harmonic geometric structures (including analytic $\mathcal{N}=2$ gauge prepotentials) should underlie $\mathcal{N}=2$ AdS supergravity as well.

\smallskip

$\bullet$  \textit{$\mathcal{N}=2$ AdS higher spins }

\smallskip

Analogously to the Minkowski case, we expect that the analytic prepotentials of linearized $\mathcal{N}=2$ AdS
supergravity have simple harmonic transformation laws and admit a straightforward generalization to higher-spin theories.
The approach we have developed will presumably allow one to construct such prepotentials by gauging rigid AdS supersymmetries
realized on the hypermultiplet, analogously to the flat super Minkowski case (see \cite{Buchbinder:2025ceg} for a recent review).
We hope to move on along this direction in the future study.

\smallskip

Long time ago, $\mathcal{N}=2$ higher spin theories were constructed in terms of $\mathcal{N}=1$ superfields
by Gates, Kuzenko and Sibiryakov \cite{Gates:1996my, Gates:1996xs}, and then manifestly $OSp(2|4)$  invariant equations of motion for such theories were constructed
by Segal and Sibiryakov \cite{Segal:1999qr} without using HSS methods.
However, the  results obtained did not allow to construct a manifestly $OSp(2|4)$ invariant action.
A recent construction of $\mathcal{N}=2$ higher spins in flat harmonic superspace \cite{Buchbinder:2021ite}
implies the obvious opportunity of generalizing  the seminal results mentioned above to $\mathcal{N}=2$ AdS and constructing a manifestly $OSp(2|4)$ invariant superspace action.

\acknowledgments

%\vspace{-0.2 cm}

The present work  was supported by the Foundation for the Advancement of Theoretical Physics and
Mathematics ``BASIS'',  grant \verb|#| 25-1-1-10-4.

\appendix

\section{Some useful relations}\label{eq: App A}
It is convenient to use the notation
\be
y := c^{+-}\,, \quad z := x^2 = x^{\alpha\dot{\alpha}} x_{\alpha\dot{\alpha}}, \, \quad \epsilon^{\pm\alpha} = \epsilon^{\alpha i}u^{\pm}_i\,, \;
\bar{\epsilon}^{\pm\dot\alpha} = \bar{\epsilon}^{\dot\alpha i}u^{\pm}_i  = \overline{(\epsilon^{\pm\alpha})} \lb{defzy}
\ee
and the relations
\bea
&&\eta^{+\alpha} = c^{ik}u^+_i \epsilon_{k}^\alpha = c^{++} \epsilon^{-\alpha} - y\epsilon^{+\alpha}\,,
\quad
\eta^{-\alpha}  = c^{ik}u^-_i \epsilon_{k}^\alpha = -c^{--} \epsilon^{+\alpha} + y\epsilon^{-\alpha}\,,\lb{cue1} \\
&& D^{++} c^{++} = 0\,, \quad D^{++} y = c^{++}\,, \quad D^{++} c^{--} =  2y\,, \quad c^{++}c^{--} = y^2 + m^2\,, \; m^2 := \frac{c^2}{2}\,. \lb{cue2}
\eea
Over the paper we make use the ``bar'' (or ``overline'') symbol to denote both the ordinary complex conjugation (on the ordinary coordinates and fields) and the ``tilde''
conjugation which is pertinent just to the harmonic variables and is squared to $-1$. In particular,
\bea
\overline{(u^\pm_i)} = u^{\pm i}\,, \quad \overline{(u^{\pm i})} = -  u^{\pm}_{ i}\,.
\eea
The relevant conjugation properties of other quantities in HSS are summarized in the Appendix of the book \cite{18}.

It is worth manifestly giving the corresponding holomorphic parts of the transformations of various Lorentz covariant expressions:
\begin{equation}
	\begin{split}
		& \delta_{\epsilon} z = \delta_{\epsilon} x^2 =
		- 8i x_{\alpha\dot\alpha} \epsilon^{-\alpha}\bar\theta^{+\dot\alpha} - 4i z \theta^{+\alpha} (y\epsilon^-_\alpha  - c^{--}\epsilon^+_\alpha )\,, \; \\
		& \delta_{\epsilon} y = 4i (y^2 + m^2)(\theta^+\epsilon^-) - 4i y c^{--}(\theta^+\epsilon^+),\\
		& \delta_{\epsilon}(\theta^+)^2 = 2 (\theta^+\epsilon^{+})\,, \quad \delta_\epsilon (\bar\theta^+)^2 = -2 x^{\alpha\dot{\alpha}} \eta^+_\alpha \bar{\theta}^+_{\dot{\alpha}} = -2x^{\alpha\dot\alpha}  (c^{++}\epsilon^-_\alpha - y \epsilon^+_\alpha)\bar\theta^+_{\dot\alpha}\,,
		\\
		& \delta_{\epsilon} (\theta^{+\alpha}\bar\theta^{+\dot\alpha}) = \epsilon^{+\alpha} \bar\theta^{+ \dot\alpha}
		+ 2i\,(\theta^+)^2 \bar\theta^{+\dot\alpha} (y\epsilon^-_\alpha  - c^{--}\epsilon^+_\alpha ) + x^{\beta\dot\alpha} \theta^{+ \alpha}(c^{++}\epsilon^-_\alpha - y \epsilon^+_\alpha)\,,% \lb{TranCovar1}
		\\
		& \delta_{\epsilon} (x_{\alpha\dot\alpha}\theta^{+\alpha}\bar\theta^{+\dot\alpha}) = -(\theta^+\epsilon^-) \big[4i(\bar\theta^+)^2 - \frac{z}{2}c^{++}\big] - (\theta^+\epsilon^+) \frac12 zy
		+ 2iy(x_{\alpha\dot\alpha}\epsilon^{-\alpha}\bar\theta^{+\dot\alpha})(\theta^+)^2 \\
		& \qquad \qquad \qquad \quad\;\;\;\; +\, (x_{\alpha\dot\alpha}\epsilon^{+\alpha}\bar\theta^{+\dot\alpha})\big[1 - 2ic^{--} (\theta^+)^2\big].\lb{TranCovar2}
	\end{split}
\end{equation}

Also, under the $SO(2)$ transformations \eqref{U1}, the quantities $z$, $y$ and $c^{++}$ transform as
\bea
\delta_{\gamma} z = -8i\gamma
c^{--} x_{\alpha\dot\alpha} \theta^{+\alpha}\bar\theta^{+\dot\alpha}\,, \quad \delta_{\gamma} y = \gamma (y^2 + m^2)\,, \quad  \delta_{\gamma} c^{++} = 2 \gamma y c^{++}\,. \lb{TranU1}
\eea
It is worth explicitly giving the nonlinear translation transformations of these objects
\bea
\delta_a z = 2(ax)\left(1 + \frac{m^2}{2} z\right)\,, \;\;
\delta_a y = 4i m^2 c^{--}a_{\alpha\dot\alpha} \theta^{+\alpha}\bar\theta^{+ \dot\alpha}\,, \;\;  \delta_a c^{++} =
8 i m^2 y \,a_{\alpha\dot\alpha}\theta^{+\alpha}\bar\theta^{+\dot\alpha}\,. \lb{aTranzy}
\eea

Using the transformation law $\delta_a z$ we deduce the transformation law for $e(x)^n$:
\begin{equation}\label{eq: a^n}
	\delta_a e(x)^n
	=
	\delta_a \left(1+ \frac{m^2}{2}z \right)^{-n} =
	-
	n m^2 (ax) \left(1+ \frac{m^2}{2}z \right)^{-n}
	=
	- n m^2(ax) \, e(x)^n.
\end{equation}

Note also some useful corollaries of the $SO(2)$ transformation law of $y$ in \eqref{TranU1}:
\bea
\delta_{\gamma} \left(\frac{y^2}{m^2} + 1\right)^n  = 2n\gamma\,y\,\left(\frac{y^2}{m^2} + 1\right)^n\,, \quad
\delta_{\gamma} \log\left(\frac{y^2}{m^2} + 1\right) = 2\gamma\,y\,.\lb{U14}
\eea

\section{Redefinition of Grassmann coordinates}\label{app B}

The ordinary derivative $\partial_{\alpha\dot\beta}$ is not covariant under non-linear AdS translations  \eqref{NonlTran}:
\begin{equation}
    \delta_a \partial_{\alpha\dot{\beta}} = - m^2 (ax) \partial_{\alpha\dot{\beta}}
    +
    \ell_{(\alpha\gamma)} \partial^\gamma_{\;\dot\beta}
    +  \bar{\ell}_{(\dot\alpha\dot\gamma)}\partial_\alpha^{\;\dot\gamma},
\end{equation}
where
\be\label{eq: nonl Lorentz}
\ell_{(\alpha\gamma)} := \frac{c^2}{2}\,x^{\;\dot\beta}_{(\alpha}a_{\gamma) \dot\beta}\,
\quad
\bar{\ell}_{(\dot\alpha\dot\gamma)} := \frac{c^2}{2}\,x^{\beta}_{\;(\dot\alpha}a_{\beta \dot\gamma)}\,,
\ee
are parameters of the nonlinearly realized Lorentz group.

The object
\begin{equation}
\nabla_{\alpha\dot\beta} := \big(1 + \frac{m^2}{2} x^2\big)\partial_{\alpha\dot\beta}
=
e(x)^{-1} \partial_{\alpha\dot{\alpha}} \lb{AdScovx}
\end{equation}
is transformed as
\begin{equation}
\delta_a \nabla_{\alpha\dot\beta} =
\ell_{(\alpha\gamma)} \nabla^\gamma_{\;\dot\beta}
+  \bar{\ell}_{(\dot\alpha\dot\gamma)}\nabla_\alpha^{\;\dot\gamma}  , \lb{LorPar}
\end{equation}
So under nonlinear AdS translations $\nabla_{\alpha\dot\beta}$ undergoes, in their spinorial indices,
just induced Lorentz rotations and so is the sought $SO(3,2)$ covariant derivative with respect to $x^{\alpha\dot\alpha}$. There still remains the problem
of defining the proper covariant spinor derivative possessing the correct transformation properties under Lorentz rotations
with the parameters \eqref{LorPar}.

In order to gain the $x$-covariant derivative in $D^{++}$, we need to make the redefinition
\begin{equation}
\big( \theta^{+\alpha}, \bar\theta^{+\dot\alpha} \big) =  e(x)^{-\frac{1}{2}}\big( \hat{\theta}^{+\alpha}, \hat{\bar{\theta}}^{+\dot\alpha} \big),\lb{RederGr}
\end{equation}
whence
\begin{equation}
D^{++} = \partial^{++} -4i\hat{\theta}^{+\alpha}\hat{\bar{\theta}}^{+\dot\alpha}\nabla_{\alpha\dot\alpha}
- im^2 \Big((\hat{\theta}^+)^2 x^\beta_{\dot\alpha}\hat{\bar{\theta}}^{+ \dot\alpha} \frac{\partial}
{\partial\hat{\theta}^{+\beta}} + (\hat{\bar{\theta}}^{+})^2  x_\alpha^{\dot\beta}\hat{\theta}^{+\alpha} \frac{\partial}{\partial\hat{\bar{\theta}}^{+\dot\beta}} \Big).
\end{equation}
Also, the analytic superspace integration measure should undergo the additional transformation
\begin{equation}
d\zeta^{(-4)}  =  d\hat{\zeta}^{(-4)}\, {\rm Ber}\,\left(\frac{\partial \zeta}{\partial \hat{\zeta}}\right)
=
 d\hat{\zeta}^{(-4)} e(x)^2
\,. \lb{Berez}
\end{equation}

Note that under AdS translations the newly defined Grassmann variables  have the same transformation laws as $\nabla_{\alpha\dot\alpha}$:
\begin{equation}
\delta_{a} \hat{\theta}^{+\alpha} = -\ell^{(\alpha\beta)}\hat{\theta}^{+}_{\beta}\,, \quad \delta_{a} \hat{\bar{\theta}}^{+\dot\alpha} =  -\bar{\ell}^{(\dot\alpha\dot\beta)}\hat{\bar{\theta}}^{+}_{\dot\beta}\,.\lb{NonlLor1}
\end{equation}
However, in contrast to the original coordinates $\theta^+_\alpha, \bar{\theta}^+_{\dot\alpha}$, the coordinates $\hat{\theta}^+_\alpha, \hat{\bar{\theta}}^+_{\dot\alpha}$
have more  complicated $SO(2)$ (and AdS supersymmetry) transformation laws.
So it is natural to attempt a more general redefinition of the original Grassmann coordinates  ${\theta}^{+}_{\alpha}, {\bar{\theta}}^{+}_{\dot\alpha}$
\bea
&& \hat{\theta}^{+ \alpha} = f_0 (z, y)\,{\theta}^{+ \alpha} + c^{--} f_1(z, y)\,(\bar\theta^+)^2 \theta^{ + \alpha} + c^{--} f_2(z, y)\,(\theta^+)^2 x^{\alpha\,\,}_{\dot\alpha}\bar\theta^{+\dot\alpha}\,, \lb{GenRed}
\eea
and to require new coordinates to transform under the nonlinear translations as in \eqref{NonlLor1}
\begin{subequations}\label{InvProp}
	\bea
	\delta_{a} \hat{\theta}^{+\alpha}
	= -\ell^{(\alpha\beta)}\hat{\theta}^{+}_{\beta}\, \lb{InvProp a}
	\eea
	and at the same time to have  simpler $SO(2)$ transformation law
	\bea
	\delta_\gamma \hat{\theta}^{+ \alpha} = \lambda \gamma y\,  \hat{\theta}^{+\alpha}\,, \lb{InvProp b}
	\eea
\end{subequations}
with $\lambda$ some constant.

The coefficients in \eqref{GenRed}, before imposing the conditions \eqref{InvProp}, are some complex functions
and we are interested in the non-singular solutions,  $f_0 = 1 + \ldots$.
The condition \eqref{InvProp b}  amounts to the following equations:
\begin{equation}
	\begin{split}
		\theta^{+\alpha}: \qquad & (a)\; (y^2 + m^2)\partial_y f_0 + y f_0 =  \lambda y f_0\,,
		\\
		(\bar{\theta}^+)^2 \theta^{+\alpha} c^{--}: \qquad&  (b)\;(y^2 + m^2)\partial_y f_1 + 3 y f_1 =  \lambda y f_1\,,
		\\
		(\theta^+)^2 \bar{\theta}^{+\dot{\beta}} x^\alpha_{\dot{\beta}} c^{--}:
		\qquad
		&  (c)\;(y^2 + m^2)\partial_y f_2 + 3 y f_2 + 4i \partial_z f_0 = \lambda y f_2\,. \lb{gammaEqs}
	\end{split}
\end{equation}
Equation \eqref{InvProp a}   gives:
\begin{equation} \label{aEqs}
	\begin{split}
		\theta^{+\alpha}:
		\qquad
		& (a)\;(1 + \frac{m^2}{2} z)\partial_z f_0 + \frac{m^2}{4} f_0 = 0\,,
		\\
		(\bar{\theta}^+)^2 \theta^{+\alpha}:
		\qquad
		& (b)\;(1 + \frac{m^2}{2} z)\partial_z f_1 + \frac{3m^2}{4} f_1 = 0\,,
		\\
		(\theta^+)^2 x^\alpha_{\dot\alpha} \bar{\theta}^{\dot\alpha} c^{--}:
		\qquad
		& (c) \; (1 + \frac{m^2}{2} z)\partial_z f_2 + \frac{3m^2}{4} f_2 = 0\,,
		\\
		(\theta^+)^2 \bar{\theta}^{+\dot\alpha} a^\alpha_{\dot\alpha} c^{--}:
		\qquad
		& (d)\; (1 + \frac{m^2}{2} z)f_2 - 2im^2\partial_y f_0  = 0\,.
	\end{split}
\end{equation}
The appropriate solution surprisingly exists only for  $\lambda = \frac{3}{2}$:
\begin{equation}
f_0 = e(x)^{\frac{1}{2}}\left( 1 + \frac{y^2}{m^2}\right)^{\frac{1}{4}}\,,
\;\;
\;
f_1 = \beta_1 e(x)^{\frac{3}{2}}\left( 1 + \frac{y^2}{m^2}\right)^{-\frac{3}{2}}\,,
\;\;
\; f_2 =  iye(x)^{\frac{3}{2}}\  \left( 1 + \frac{y^2}{m^2}\right)^{-\frac{3}{4}}\,. \lb{Solut}
\end{equation}
The integration constant $\beta_1$ can be chosen zero, so the $f_1$ term in \eqref{GenRed} is unessential, while the $f_2$ term
is essential for retrieving a non-singular solution for $f_0$. The geometric meaning of such coordinate redefinition is not clear for us
for the moment. In principle, when performing the component calculations, one is not obliged to care about the $\gamma$ transformation properties
of Grassmann coordinates and can stick to the simplest coordinate change \eqref{RederGr} - \eqref{Berez}.

\vspace{-0.4 cm}
%\newpage


\begin{thebibliography}{99}

%\cite{Galperin:1984av}
\bibitem{Galperin:1984av}
A.~Galperin, E.~Ivanov, S.~Kalitzin, V.~Ogievetsky and E.~Sokatchev,
{\it Unconstrained $\mathcal{N}=2$ Matter, Yang-Mills and Supergravity Theories in Harmonic Superspace},
\href{https://doi.org/10.1088/0264-9381/1/5/004}{Class. Quant. Grav. \textbf{1} (1984), 469-498}
[erratum: \href{https://doi.org/10.1088/0264-9381/2/1/512}{Class. Quant. Grav. \textbf{2} (1985), 127}].
%848 citations counted in INSPIRE as of 14 Nov 2024

\bibitem{18} A.~S.~Galperin, E.~A.~Ivanov, V.~I.~Ogievetsky, E.~S.~Sokatchev,
{\it Harmonic superspace}, Cambridge Monographs on Mathematical
Physics, Cambridge University Press, 2001, 306 p.

%\cite{Galperin:1987em}
\bibitem{Galperin:1987em}
A.~S.~Galperin, N.~A.~Ky and E.~Sokatchev,
{\it $\mathcal{N}=2$ Supergravity in Superspace: Solution to the Constraints},
\href{https://doi.org/10.1088/0264-9381/4/5/022}{Class. Quant. Grav. \textbf{4} (1987), 1235}.
%55 citations counted in INSPIRE as of 14 Nov 2024

%\cite{Galperin:1987ek}
\bibitem{Galperin:1987ek}
A.~S.~Galperin, E.~A.~Ivanov, V.~I.~Ogievetsky and E.~Sokatchev,
{\it $\mathcal{N}=2$ Supergravity in Superspace: Different Versions and Matter Couplings},
\href{https://doi.org/10.1088/0264-9381/4/5/023}{Class. Quant. Grav. \textbf{4} (1987), 1255}.
%96 citations counted in INSPIRE as of 14 Nov 2024


%\cite{Buchbinder:2021ite}
\bibitem{Buchbinder:2021ite}
I.~Buchbinder, E.~Ivanov and N.~Zaigraev,
{\it Unconstrained off-shell superfield formulation of $4D,  \mathcal{N}=2$ supersymmetric higher spins},
\href{https://doi.org/10.1007/JHEP12(2021)016}{JHEP \textbf{12} (2021), 016}
[arXiv:2109.07639 [hep-th]].
%19 citations counted in INSPIRE as of 14 Nov 2024

%\cite{Buchbinder:2022kzl}
\bibitem{Buchbinder:2022kzl}
I.~Buchbinder, E.~Ivanov and N.~Zaigraev,
{\it Off-shell cubic hypermultiplet couplings to $\mathcal{N}=2$ higher spin gauge superfields},
\href{https://doi.org/10.1007/JHEP05(2022)104}{JHEP \textbf{05} (2022), 104}
[arXiv:2202.08196 [hep-th]].
%17 citations counted in INSPIRE as of 14 Nov 2024


%\cite{Buchbinder:2022vra}
\bibitem{Buchbinder:2022vra}
I.~Buchbinder, E.~Ivanov and N.~Zaigraev,
{\it $\mathcal{N}=2$ higher spins: superfield equations of motion, the hypermultiplet supercurrents, and the component structure},
\href{https://doi.org/10.1007/JHEP03(2023)036}{JHEP \textbf{03} (2023), 036}
[arXiv:2212.14114 [hep-th]].
%9 citations counted in INSPIRE as of 14 Nov 2024





%\cite{Buchbinder:2024pjm}
\bibitem{Buchbinder:2024pjm}
I.~Buchbinder, E.~Ivanov and N.~Zaigraev,
{\it $\mathcal{N}=2$ superconformal higher-spin multiplets and their hypermultiplet couplings},
\href{https://doi.org/10.1007/JHEP08(2024)120}{JHEP \textbf{08} (2024), 120}
[arXiv:2404.19016 [hep-th]].
%4 citations counted in INSPIRE as of 14 Nov 2024

%\cite{Buchbinder:2024xll}
\bibitem{Buchbinder:2024xll}
I.~Buchbinder, E.~Ivanov and N.~Zaigraev,
{\it $\mathcal{N} = 2$ higher spin theories and harmonic superspace},
\href{https://doi.org/10.22323/1.455.0048}{PoS \textbf{ICPPCRubakov2023} (2024), 048}
[arXiv:2402.05704 [hep-th]].
%2 citations counted in INSPIRE as of 16 Aug 2025

%\cite{Buchbinder:2025ceg}
\bibitem{Buchbinder:2025ceg}
I.~Buchbinder, E.~Ivanov and N.~Zaigraev,
{\it Towards $\mathcal{N}=2$ higher-spin supergravity},
[arXiv:2503.02438 [hep-th]].
%1 citations counted in INSPIRE as of 27 Jul 2025

%\cite{Kuzenko:1994dm}
\bibitem{Kuzenko:1994dm}
S.~M.~Kuzenko and A.~G.~Sibiryakov,
{\it Free massless higher superspin superfields on the anti-de Sitter superspace},
Phys. Atom. Nucl. \textbf{57} (1994), 1257-1267
[arXiv:1112.4612 [hep-th]].
%84 citations counted in INSPIRE as of 16 Aug 2025

%\cite{Gates:1996my}
\bibitem{Gates:1996my}
S.~J.~Gates, Jr., S.~M.~Kuzenko and A.~G.~Sibiryakov,
{\it $\mathcal{N}=2$ supersymmetry of higher superspin massless theories},
\href{https://doi.org/10.1016/S0370-2693(97)01037-X}{Phys. Lett. B \textbf{412} (1997), 59-68}
[arXiv:hep-th/9609141 [hep-th]].
%53 citations counted in INSPIRE as of 04 Aug 2025

%\cite{Gates:1996xs}
\bibitem{Gates:1996xs}
S.~J.~Gates, Jr., S.~M.~Kuzenko and A.~G.~Sibiryakov,
{\it Towards a unified theory of massless superfields of all superspins},
\href{https://doi.org/10.1016/S0370-2693(97)00034-8}{Phys. Lett. B \textbf{394} (1997), 343-353}
[arXiv:hep-th/9611193 [hep-th]].
%52 citations counted in INSPIRE as of 20 Aug 2025

%\cite{Segal:1999qr}
\bibitem{Segal:1999qr}
A.~Y.~Segal and A.~G.~Sibiryakov,
{\it Explicit $\mathcal{N}=2$ supersymmetry for higher spin massless fields in $D = 4$ AdS superspace},
\href{https://doi.org/10.1142/S0217751X02006195}{Int. J. Mod. Phys. A \textbf{17} (2002), 1207-1254}
[arXiv:hep-th/9903122 [hep-th]].
%10 citations counted in INSPIRE as of 04 Aug 2025



%\cite{Gates:1983nr}
\bibitem{Gates:1983nr}
S.~J.~Gates, M.~T.~Grisaru, M.~Rocek and W.~Siegel,
{\it Superspace Or One Thousand and One Lessons in Supersymmetry},
Front. Phys. \textbf{58} (1983), 1-548
1983,
%ISBN 978-0-8053-3161-5
[arXiv:hep-th/0108200 [hep-th]].
%955 citations counted in INSPIRE as of 28 Oct 2025

%\cite{Bandos:2002nn}
\bibitem{Bandos:2002nn}
I.~A.~Bandos, E.~Ivanov, J.~Lukierski and D.~Sorokin,
{\it On the superconformal flatness of AdS superspaces},
\href{https://doi.org/10.1088/1126-6708/2002/06/040}{JHEP \textbf{06} (2002), 040}
[arXiv:hep-th/0205104 [hep-th]].
%55 citations counted in INSPIRE as of 21 Jul 2025


%\cite{Galperin:1985zv}
\bibitem{Galperin:1985zv}
A.~Galperin, E.~Ivanov, V.~Ogievetsky and E.~Sokatchev,
{\it Conformal invariance in harmonic superspace},
JINR-E2-85-363.
%2 citations counted in INSPIRE as of 08 Jan 2023

%\cite{Kuzenko:2007aj}
\bibitem{Kuzenko:2007aj}
S.~M.~Kuzenko and G.~Tartaglino-Mazzucchelli,
{\it Five-dimensional $\mathcal{N} = 1$ AdS superspace: Geometry, off-shell multiplets and dynamics},
\href{https://doi.org/10.1016/j.nuclphysb.2007.06.014}{Nucl. Phys. B \textbf{785} (2007), 34-73}
[arXiv:0704.1185 [hep-th]].
%39 citations counted in INSPIRE as of 04 Aug 2025

%\cite{Kuzenko:2007vs}
\bibitem{Kuzenko:2007vs}
S.~M.~Kuzenko and G.~Tartaglino-Mazzucchelli,
{\it On 5D AdS SUSY and harmonic superspace}, in  7th International Workshop on Supersymmetries and Quantum Symmetries (SQS'07),
[arXiv:0711.0063 [hep-th]].
%3 citations counted in INSPIRE as of 16 Aug 2025

\bibitem{Kuzenko:2008qw}
S.~M.~Kuzenko and G.~Tartaglino-Mazzucchelli,
{\it Field theory in $4D \; \mathcal{N}=2$ conformally flat superspace},
\href{https://doi.org/10.1088/1126-6708/2008/10/001}{JHEP \textbf{10} (2008), 001}
[arXiv:0807.3368 [hep-th]].

%\cite{Ivanov:1980vb}
\bibitem{Ivanov:1980vb}
E.~A.~Ivanov and A.~S.~Sorin,
{\it Superfield formulation of $OSp(1,4)$ supersymmetry},
\href{https://doi.org/10.1088/0305-4470/13/4/013}{J. Phys. A \textbf{13} (1980), 1159-1188}.
%115 citations counted in INSPIRE as of 02 Aug 2025


%\cite{deWit:1987sn}
\bibitem{deWit:1987sn}
B.~de Wit and A.~Zwartkruis,
{\it $SU(2,2|1,1)$ Supergravity and $\mathcal{N}=2$ Supersymmetry With Arbitrary Cosmological Constant},
\href{https://doi.org/10.1088/0264-9381/4/3/005}{Class. Quant. Grav. \textbf{4} (1987), L59}.
%12 citations counted in INSPIRE as of 22 Aug 2025

%\cite{Ivanov:2013cea}
\bibitem{Ivanov:2013cea}
E.~Ivanov and S.~Sidorov,
{\it Super K{\"a}hler oscillator from $SU(2|1)$ superspace},
\href{https://doi.org/10.1088/1751-8113/47/29/292002}{J. Phys. A \textbf{47} (2014), 292002}
[arXiv:1312.6821 [hep-th]].
%35 citations counted in INSPIRE as of 02 Oct 2025

%\cite{Ivanov:2015iia}
\bibitem{Ivanov:2015iia}
E.~Ivanov, S.~Sidorov and F.~Toppan,
{\it Superconformal mechanics in $SU(2|1)$ superspace},
\href{https://doi.org/10.1103/PhysRevD.91.085032}{Phys. Rev. D \textbf{91} (2015) no.8, 085032}
[arXiv:1501.05622 [hep-th]].
%31 citations counted in INSPIRE as of 02 Oct 2025


%\cite{Lukierski:1984it}
\bibitem{Lukierski:1984it}
J.~Lukierski and A.~Nowicki,
{\it All Possible De Sitter Superalgebras and the Presence of Ghosts},
\href{https://doi.org/10.1016/0370-2693(85)91659-4}{Phys. Lett. B \textbf{151} (1985), 382-386}.
%50 citations counted in INSPIRE as of 22 Aug 2025

%\cite{Pilch:1984aw}
\bibitem{Pilch:1984aw}
K.~Pilch, P.~van Nieuwenhuizen and M.~F.~Sohnius,
{\it De Sitter Superalgebras and Supergravity},
\href{https://doi.org/10.1007/BF01211046}{Commun. Math. Phys. \textbf{98} (1985), 105}.
%119 citations counted in INSPIRE as of 22 Aug 2025

%\cite{Ohta:1985ba}
\bibitem{Ohta:1985ba}
N.~Ohta, H.~Sugata and H.~Yamaguchi,
{\it $\mathcal{N}=2$ Harmonic Superspace With Central Charges and Its Application to Selfinteracting Massive Hypermultiplets},
\href{https://doi.org/10.1016/0003-4916(86)90018-7}{Annals Phys. \textbf{172} (1986), 26}.
%16 citations counted in INSPIRE as of 30 Sep 2025

%\cite{Buchbinder:1997pw}
\bibitem{Buchbinder:1997pw}
I.~L.~Buchbinder, E.~I.~Buchbinder, E.~A.~Ivanov, S.~M.~Kuzenko and B.~A.~Ovrut,
{\it Effective action of the $\mathcal{N}=2$ Maxwell multiplet in harmonic superspace},
\href{https://doi.org/10.1016/S0370-2693(97)01025-3}{Phys. Lett. B \textbf{412} (1997), 309-319}
[arXiv:hep-th/9703147 [hep-th]].
%70 citations counted in INSPIRE as of 29 Sep 2025


%\cite{Ivanov:2024gjo}
\bibitem{Ivanov:2024gjo}
E.~Ivanov and N.~Zaigraev,
{\it Off-shell invariants of linearized $4D,\mathcal{N}=2$ supergravity in the harmonic approach},
\href{ttps://doi.org/10.1103/PhysRevD.110.066020}{Phys. Rev. D \textbf{110} (2024) no.6, 066020}
[arXiv:2407.08524 [hep-th]].
%3 citations counted in INSPIRE as of 30 Sep 2025

%\cite{Freedman:1983na}
\bibitem{Freedman:1983na}
D.~Z.~Freedman and H.~Nicolai,
{\it Multiplet Shortening in $OSp(N,4)$},
\href{https://doi.org/10.1016/0550-3213(84)90164-0}{Nucl. Phys. B \textbf{237} (1984), 342-366}.
%102 citations counted in INSPIRE as of 18 Aug 2025

%\cite{Breitenlohner:1982bm}
\bibitem{Breitenlohner:1982bm}
P.~Breitenlohner and D.~Z.~Freedman,
{\it Positive Energy in anti-De Sitter Backgrounds and Gauged Extended Supergravity},
\href{https://doi.org/10.1016/0370-2693(82)90643-8}{Phys. Lett. B \textbf{115} (1982), 197-201}.
%1301 citations counted in INSPIRE as of 05 Aug 2025

\bibitem{IvSor2}
E.A.~Ivanov and A.S.~Sorin,{\it Wess-Zumino model as linear sigma model of spontaneously broken conformal and $OSp(1,4)$ supersymmetries},
Yad. Fiz. {\bf 30} (1979) 853-866 [Sov. J. Nucl. Phys. {\bf 30} (1979) 440].


%\cite{Galperin:1985de}
\bibitem{Galperin:1985de}
A.~Galperin, E.~Ivanov, V.~Ogievetsky and E.~Sokatchev,
{\it Hyperkahler Metrics and Harmonic Superspace},
\href{https://doi.org/10.1007/BF01211764}{Commun. Math. Phys. \textbf{103} (1986), 515}.
%96 citations counted in INSPIRE as of 03 Aug 2025

%\cite{Bagger:1987rc}
\bibitem{Bagger:1987rc}
J.~A.~Bagger, A.~S.~Galperin, E.~A.~Ivanov and V.~I.~Ogievetsky,
{\it Gauging $\mathcal{N}=2$ Sigma Models in Harmonic Superspace},
\href{https://doi.org/10.1016/0550-3213(88)90392-6}{Nucl. Phys. B \textbf{303} (1988), 522-542}.
%61 citations counted in INSPIRE as of 01 Oct 2025


%\cite{Ivanov:2022vwc}
\bibitem{Ivanov:2022vwc}
E.~Ivanov,
{\it $\mathcal{N}\,{=}\,2$ Supergravities in Harmonic Superspace}, in: \href{https://doi.org/10.1007/978-981-19-3079-9\textunderscore43-1}{Bambi, C., Modesto, L., Shapiro, I. (eds) Handbook of Quantum Gravity. Springer, Singapore}.
[arXiv:2212.07925 [hep-th]].
%8 citations counted in INSPIRE as of 18 Aug 2025

%\cite{Butter:2015nza}
\bibitem{Butter:2015nza}
D.~Butter,
{\it On conformal supergravity and harmonic superspace},
\href{https://doi.org/10.1007/JHEP03(2016)107}{JHEP \textbf{03} (2016), 107}
[arXiv:1508.07718 [hep-th]].
%13 citations counted in INSPIRE as of 18 Aug 2025

%\cite{Koning:2024iiq}
\bibitem{Koning:2024iiq}
N.~E.~Koning, S.~M.~Kuzenko and E.~S.~N.~Raptakis,
{\it The anti-de Sitter supergeometry revisited},
\href{https://doi.org/110.1007/JHEP02(2025)175}{JHEP \textbf{02} (2025), 175}
[arXiv:2412.03172 [hep-th]].
%5 citations counted in INSPIRE as of 04 Aug 2025




%\cite{Galperin:1992pj}
\bibitem{Galperin:1992pj}
A.~Galperin, E.~Ivanov and O.~Ogievetsky,
{\it Harmonic space and quaternionic manifolds},
\href{https://doi.org/10.1006/aphy.1994.1025}{Annals Phys. \textbf{230} (1994), 201-249}
[arXiv:hep-th/9212155 [hep-th]].
%49 citations counted in INSPIRE as of 18 Aug 2025





%\cite{Kuzenko:2008qw}

%50 citations counted in INSPIRE as of 04 Aug 2025



%\cite{Butter:2011zt}
\bibitem{Butter:2011zt}
D.~Butter and S.~M.~Kuzenko,
{\it $\mathcal{N}=2$ supersymmetric sigma-models in AdS},
\href{https://doi.org/10.1016/j.physletb.2011.08.043}{Phys. Lett. B \textbf{703} (2011), 620-626}
[arXiv:1105.3111 [hep-th]].
%34 citations counted in INSPIRE as of 04 Aug 2025

%\cite{Butter:2011kf}
\bibitem{Butter:2011kf}
D.~Butter and S.~M.~Kuzenko,
{\it The structure of $\mathcal{N}=2$ supersymmetric nonlinear sigma models in AdS$_4$},
\href{https://doi.org/10.1007/JHEP11(2011)080}{JHEP \textbf{11} (2011), 080}
[arXiv:1108.5290 [hep-th]].
%37 citations counted in INSPIRE as of 29 Aug 2025

%\cite{Butter:2012jj}
\bibitem{Butter:2012jj}
D.~Butter, S.~M.~Kuzenko, U.~Lindstrom and G.~Tartaglino-Mazzucchelli,
{\it Extended supersymmetric sigma models in AdS$_4$ from projective superspace},
\href{https://doi.org/10.1007/JHEP05(2012)138}{JHEP \textbf{05} (2012), 138}
[arXiv:1203.5001 [hep-th]].
%21 citations counted in INSPIRE as of 29 Aug 2025

\bibitem{Sergei}
S.M.~Kuzenko, {\it Projective superspace as a double-pinctured harmonic superspace}, \href{https://doi.org/10.1142/S0217751X99000889}{Int. J. Mod. Phys. A{\bf 14} (1999) 1737} [arXiv:hep-th/9806147].

%\cite{Butter:2011ym}
\bibitem{Butter:2011ym}
D.~Butter and S.~M.~Kuzenko,
{\it $\mathcal{N}=2$ AdS supergravity and supercurrents},
\href{https://doi.org/10.1007/JHEP07(2011)081}{JHEP \textbf{07} (2011), 081}
[arXiv:1104.2153 [hep-th]].
%33 citations counted in INSPIRE as of 04 Aug 2025

%\cite{Butter:2010jm}
\bibitem{Butter:2010jm}
D.~Butter and S.~M.~Kuzenko,
{\it New higher-derivative couplings in $4D\, \mathcal{N} = 2$ supergravity},
\href{https://doi.org/10.1007/JHEP03(2011)047}{JHEP \textbf{03} (2011), 047}
[arXiv:1012.5153 [hep-th]].
%48 citations counted in INSPIRE as of 01 Oct 2025

%\cite{Butter:2010sc}
\bibitem{Butter:2010sc}
D.~Butter and S.~M.~Kuzenko,
{\it $\mathcal{N}=2$ supergravity and supercurrents},
\href{https://doi.org/10.1007/JHEP12(2010)080}{JHEP \textbf{12} (2010), 080}
[arXiv:1011.0339 [hep-th]].
%32 citations counted in INSPIRE as of 20 Aug 2025



%\cite{Buchbinder:1998qv}
%\bibitem{Buchbinder:1998qv}
%I.~L.~Buchbinder and S.~M.~Kuzenko,
%{\it Ideas and methods of supersymmetry and supergravity: Or a walk through superspace},
%Taylor and Francis, 1998.
%57 citations counted in INSPIRE as of 27 Feb 2024

\end{thebibliography}
\end{document}